\documentclass[10pt]{article}

\usepackage{lscape}
\usepackage{natbib}
\usepackage{amssymb}
\usepackage{amsfonts}
\usepackage{amsmath}
\usepackage[english,german]{babel}
\usepackage{latexsym}
\usepackage{color}
\usepackage[pdftex]{graphicx}
\usepackage{subfigure}
\usepackage{xypic}

\newtheorem{thm}{Theorem}[section]
\newtheorem{proposition}[thm]{Proposition}

\newtheorem{lemma}[thm]{Lemma}
\newtheorem{definition}[thm]{Definition}
\newtheorem{remark}[thm]{Remark}
\newtheorem{example}[thm]{Example}

\newenvironment{Proof}{\textsc{Proof.}}{\mbox{ } \hfill $\Box$ \vspace{2mm}}

\newenvironment{rmenumerate}
    {\begin{enumerate}}
    {\end{enumerate}}

\def\dfrac{\displaystyle\frac}

\def\be{\begin{equation*}}
\def\ee{\end{equation*}}

\newcommand{\ba}{\begin{array}{ll}}
\newcommand{\bal}{\begin{array}{ll}}
\newcommand{\ea}{\end{array}}

\newcommand{\n}{\mathbb{N}}
\newcommand{\Prob}{\mathbb{P}}
\newcommand{\re}{\mathbb{R}}

\newcommand{\Ll}{\mathbb{L}}

\newcommand{\E}{\mathbb{E}}
\newcommand{\I}{{\mathcal{I}}}

\newcommand{\mm}{\mathcal{M}}

\def\O{\mathcal{O}}

\begin{document}

\selectlanguage{english}

\renewcommand{\baselinestretch}{1.2}\normalsize



\title{Pollution permits, Strategic Trading and Dynamic Technology Adoption\thanks{The authors would like to thank Matteo Bonato, Raphael Calel, Ulrich Horst, and the participants of the Humboldt Economics Seminar for their helpful discussions and comments. Financial support from the Deutsche Forschungsgemeinschaft through the SFB 649 ``Economic Risk'', the Alexander von Humboldt Foundation via a research fellowship, and from the Centre for Climate Change Economics and Policy, which is funded by the UK Economic and Social Research Council (ESRC) and Munich Re, is gratefully acknowledged. The usual disclaimers apply.}}
\author{\normalsize Santiago Moreno--Bromberg \\[8pt]
        \small  Institut f\"{u}r Mathematik   \\
        \small  Humboldt-Universit\"{a}t zu Berlin \\
        \small  Unter den Linden 6\\
        \small  10099 Berlin \\
        \small  smoreno@math.hu-berlin.de
         \and
        \normalsize  Luca Taschini \\[8pt]
        \small  Grantham Research Institute  \\
        \small  London School of Economics \\
        \small  Houghton St\\
        \small  London WC2A 2AE \\
        \small  l.taschini1@lse.ac.uk
\vspace*{0.1cm}}

\maketitle

\vspace{-.9cm}
\begin{abstract}

This paper analyzes the dynamic incentives for technology adoption under a transferable permits system, which allows for strategic trading on the permit market. Initially, firms can invest both in low--emitting production technologies and trade permits. In the model, technology adoption and allowance price are generated endogenously and are inter--dependent. It is shown that the non--cooperative permit trading game possesses a pure--strategy Nash equilibrium, where the allowance value reflects the level of uncovered pollution (demand), the level of unused allowances (supply), and the technological status. These conditions are also satisfied  when a price support instrument, which is contingent on the adoption of the new technology, is introduced. Numerical investigation confirms that this policy generates a floating price floor for the allowances, and it restores the dynamic incentives to invest. Given that this policy comes at a cost, a criterion for the selection of a \textit{self--financing} policy (based on convex risk measures) is proposed and implemented.

\end{abstract}

\vspace{0.15cm}

\begin{center}
\textsc{Preliminary - Comments Welcome}
\end{center}

\vspace{0.25cm}

\textbf{JEL classification}: D8, H2, L5, Q5.

\vspace{0.25cm}

\textbf{Keywords}: Dynamic regulation, Emission permits, Environment, Self--financing policy, Technology adoption.

\hfill

\newpage

\renewcommand{\baselinestretch}{1.0}\normalsize

\section{Introduction}

There is a wide array of pollution control instruments available to environmental regulators. Economists distinguish mainly between two types of instruments: command--and--control and market--based instruments. The most common  command--and--control instruments are technological standards and emission standards. Market--based instruments provide incentives to reduce emissions through price control, and regulated companies are free to choose their emission and abatement levels. The most commonly used market--based instruments are emission taxes and transferable permits. Under permits, in comparison to a tax schedule, a regulated firm must hold one permit for each unit of pollution it emits. When a firm is non--compliant, there is a penalty levied against it for each uncovered unit of the pollutant. Further, regulated firms can exchange unused allowances with other firms at a market price. Such a price is determined endogenously  by market mechanisms. \vspace{0.20cm}

A key consideration when choosing a policy is the incentives it provides to regulated companies to invest in new technologies or adopt alternative, low pollution--emitting technologies. Some specific types of investments (such as fuel switching in electricity production) notwithstanding, the adoption of  low pollution--emitting technologies consistently reduces further emissions. This has clear consequences on the future needs of permits and, more importantly, on the future incentives to adopt new technologies. Most of the current literature relies on calculating the aggregate cost savings achieved by regulated firms that have adopted the new technologies, but neglects the impact of aggregate reductions on the amount of unused allowances available for exchange. Moreover, such an analysis does not showcase an individual firm's incentives to adopt low pollution--emitting technology. In particular, this approach ignores the fact that some firms can free--ride on a decreasing allowance price caused by other firms' investments in abatement or low pollution--emitting technologies. The dynamic incentives to adopt new technologies  endogenously depend on the future value of the allowances, which itself depends on the future  supply and demand of permits.\footnote{Recent studies on technological change in economic models of environmental policy emphasize the need to consider technology adoption and technology innovation as endogenous decision variables rather than as exogenous processes. We refer to \cite{ELKGK:06}, \cite{Loeschel:02}, and \cite{RU:03} for further discussions. In this paper we concentrate on the incentives for the adoption of readily available low pollution--emitting technologies. Readers interested in the incentives for technology diffusion and technology innovation are referred to \cite{FPP:03}, \cite{Requate:05} and references therein.} \cite{BHQ:95} have investigated this aspect. The authors show that under a system of tradable permits, technology adoption is distorted because individual regulated companies have a significant effect on the aggregate supply of permits. They, however,  assume that companies are price--takers, and do not investigate the impact on the incentives for technology adoption of strategic exchanges of permits. As shown by \cite{KennedyLaplante:99}, under imperfect competition standard results might have to be revised. Determining the firms' optimal compliance strategy in the presence of strategic exchange of permits is part of the contributions of this paper. \vspace{0.20cm}

We first examine a pollution--constrained economy, where we assume that the regulator does not anticipate the adoption of new technology and he commits to the type of policy instrument and its level for a sufficiently long period of time. We present a relatively tractable model where regulated firms can determine their compliance strategies by choosing (not necessarily in a mutually exclusive fashion) between investment in low pollution--emitting technologies or in exchange of permits. Firms are characterized by their uncertain incomes and pollution profiles. In particular, the firms' emissions are subject to economic shocks and contingent on the types of new technologies that have been set in place. The adoption of new technologies is assumed to affect only the amount of pollution emitted for given output or input, and does not otherwise affect production. We move away from the price--taker assumption and argue that it may very well be in the sellers' best interest to strategically reduce the availability of permits and, consequently, increase the allowance exchange value. This model accounts for such strategic trading behaviors. In particular, we construct a non--cooperative permit trading game and show it possesses a pure--strategy Nash equilibrium. The expected equilibrium exchange value of permits is determined by the unique solution of the trading game. Moreover, the value of a permit reflects the economic uncertainty, the current level of uncovered pollution (demand), the current level of unused allowances (supply), and the current level of technology adoption. \vspace{0.20cm}

The incentives to invest in new technologies or adopt alternative low pollution--emitting technologies are also generated endogenously. In general, technology adoption depends on the (uncertain) future supply and demand of permits. More specifically, the incentives for a firm in permit excess hinge on the firm's potential profits, i.e.~on the ability to sell unused permits; the incentives for a firm in the need for permits depend on the firm's potentially avoided penalty costs, i.e.~on its ability to reduce emissions by the use of new technologies. An extremely low allowance price makes sales of permits unprofitable and the meeting of compliance by purchasing permits a definitely cheap alternative. A too--low allowance price, therefore, kills the incentives to adopt new technologies. The regulator may wish to intervene by adjusting the level of the policy in order to address this issue. However, the possibility of a regulator's intervention raises concerns about the time consistency of the policy, thus undermining its credibility. In light of this problem and in the spirit of \cite{LT:96b}, we implement a policy instrument that largely reduces the need of intervention by the regulator and restores the dynamic incentives to adopt low pollution--emitting technologies. The new policy consists of a price support instrument: the regulator offers each firm a fixed amount of money (contingent on the firm's technology status at a specific date) per unused allowance permit. This instrument, which we have dubbed European--Cash--4--Permits, can be considered a minimum price guarantee of sorts. \cite{BHQ:95} and \cite{KennedyLaplante:99} envisioned a similar type of policy, where the policy regulator buys back permits to adjust the supply in response to technology adoption choices. \vspace{0.20cm}

In the second part of the paper we construct the non--cooperative permit trading game in the presence of European--cash--for--permits. We show that this game possesses a pure--strategy Nash equilibrium. Moreover, we prove that the price support instrument generates a floating price floor as soon as one of the firms has adopted a new technology. Our numerical results echo the conclusions of \cite{LT:96b}. By controlling the policy level --the levels of the penalty and the price support-- we show that the regulator influences (i) the number of firms that adopt the new technologies, (ii) the timing of such adoptions. For instance, by increasing price support and accepting higher policy costs, the regulator increases the number of firms that adopt the low pollution--emitting technologies and he also induces earlier technology adoption. Evidently, the implementation of such a policy has a cost. Based on the fact that the penalty payments generate potential incomes, we define a policy to be \textit{self-financing} when tax--payers' funds are not required to cover the payments of the European--Cash--4--Permits. Within this framework, we numerically assess how likely it is (in terms of a convex risk measure) that the regulator will have to access tax-payers' funds, instead of using penalty payments.

\section{The Model}
In this section we present our model of a pollution--constrained economy under a tradeable--permits system. We assume that the environmental agency ({\sl the regulator}) does not anticipate the adoption of new technologies. The regulator chooses a credible emission reduction target, the overall length of the commitment period and the enforcement structure. We consider a dynamic and discrete--time setting, where the interval $[t, t+1]$ denotes one regulated period. Time $t=0$ represents the starting point of our analysis. The duration of the entire regulated time frame ({\sl or phase}) is $T$ periods. Regulated firms can adopt low pollution--emitting production technologies and exchange permits.\footnote{Throughout the paper we sometimes use the terms ``investment'' and ``implementation'' as an alternative to adoption.} Technology--adoption decisions are made at time $t,$ and their consequences manifest themselves at time $t+1.$ The exchange of permits takes place at time $t+1.$ The risk-averse firms are characterized by their pollution emission profiles before and after adopting new technologies, as well as by the costs of such investments. We work under the assumption that firms cannot borrow allowances against their future endowments; nor can they bank permits for future periods. Later we modify the policy by introducing a price support instrument, which will be a competitive alternative to banking. Finally, we assume the firms have access to a bank account that provides a riskless rate of return $r\ge 0,$ which for the sake of simplicity we assume to be constant over $[0, T].$


\subsection{The regulator's choice variables} \label{sec:regulator}

We consider an economy that consists of a group of polluting firms $(\I =\{1,\ldots, m\}),$ which operate under a tradable permits scheme. This system is designed, policed, monitored and enforced by the regulator, whose intention is to control pollution and  {promote} the adoption of low pollution--emitting technologies by implementing a  {credible} policy. Ideally, a sufficiently ambitious scheme should achieve the desired targets: a schedule for the allocation of permits decreasing over time sets the cap; a strict enforcement structure, a fine control for non--compliant firms, and a price attached to allowances set potential (positive and negative) payoffs from the trade of permits. Eventually, the mechanism that controls the incentive to adopt low pollution--emitting technology is determined by the potential extra profits and voided compliance costs of firms. Obviously, this incentive should depend on the allocation schedule, the penalty level and the length of the phase. \vspace{0.20cm}

Let us start by describing the allowances' schedule. It is not important for the problem at hand whether the permits are issued by auction or through some sort of grandfathering scheme, provided that the initial distribution to regulated companies does not create asymmetric market power. The regulator issues firm $i$ a number $N^i(t)$ of emission permits  at the beginning of each period, and we denote the {total per--period} cap by:
\be
     N(t):=\sum_{i\in\I} N^i(t), \quad t\in\{0,\ldots, T-1\}, \quad \mbox{and} \quad N^i(t)\in\re.
\ee
We will assume throughout this work that permits are infinitely divisible; in other words, a firm endowed with $N^i(t)$ permits may sell any real number between zero and $N^i(t)$ permits. We believe that given the very large number of unitary permits that firms receive in reality, this is quite a mild assumption, which simplifies the mathematical analysis significantly. \vspace{0.20cm}

In principle, the regulator must send the appropriate signals required to steer investors towards building a low--pollution economy. This corresponds to the identification of a sufficiently ambitious cap in terms of permitted aggregated emissions. A decreasing target $N(t)$ ought to set the desired trend of the permit price, which should ideally increase through time and should favor the adoption of more expensive technologies as time goes by and new technologies become available. In order to model the evolution of $N(t),$ while keeping our model tractable, we introduce
\be
f_{(\alpha, \beta)}(x)=\beta (x+1)^{\alpha}
\ee
and the parametric family of (non--increasing) functions:
\be
{\mathcal A}:=\big\{f_{(\alpha, \beta)}\,\mid\, \beta, -\alpha\in\n\big\}.
\ee
Then the sequences $\{N^i(t)\}=\{f_{(\alpha^i, \beta^i)}(t)\}$, for $f_{(\alpha^i, \beta^i)}\in{\mathcal A}$ represent the permit streams that may be issued by the regulator. We stress that the parametric determination of the allocation of allowances is done for computational simplicity, and does not play a role in the theoretical results that we present below. \vspace{0.20cm}

Let us now consider the second policy parameter: the penalty. In any permits scheme, there will always be a penalty for non--compliance. In our model, at the end of the $[t, t+1]$--th period, for each ton of pollution emitted that cannot be offset by an allowance, the regulated firms must pay a penalty $P.$ In this paper we assume that the penalty is an alternative to compliance, as first discussed by \cite{JacobyEllerman:04}. Consequently, the costs of technology adoption incurred by rational agents should not be greater than the expected future compliance costs. In such situations, firms would be better off paying the total penalty for uncovered future emissions rather than investing in technology adoption. In order to ignite sufficient investments in low--polluting, innovative technologies and abatement activities, therefore, the profitability of these investments needs to be fostered. In other words, the investment costs should be at least offset by those of not complying with the regulations. The regulator has no control over the investment costs, but he chooses the magnitude of the penalty. Underpricing $P$ would imply that firms preferred to pay a penalty and maintain the status quo of emissions, rather than investing to reduce their pollution footprints; whereas a prohibitively high $P$ might have a {severe} economic effect on the regulated sectors.\footnote{We refer to \cite{Cohen:99}, \cite{Keeler:91} and references therein for a comprehensive discussion about the effects of different levels of enforcement and monitoring.} \vspace{0.20cm}

Since the allowance schedule is set \textit{ex--ante}, the deterioration or improvement of the economy may encourage the regulator to adjust the level of the policy. This would clearly undermine its credibility. With the aim of reducing the need of interventions on the part of the regulator and of restoring the dynamic incentives to adopt low pollution--emitting technologies, we implement a price support instrument in Section~\ref{sec:EC4P}. This type of instrument, also investigated by \cite{LT:96b} and \cite{BHQ:95}, takes in this work the form of a free--of--charge \textit{put--type} option contract written on the final holdings of permits. More precisely, at the end of each regulated period, contingent on the adoption of the new technology, a firm can receive a pre--set amount of money, $P_{g},$ for each extra allowance. We call this instrument {\sl European--Cash--4--permits} (EC4P) and we assume that such a product is issued period--to--period, i.e.~the titles have a time to maturity of at most one period. It should be pointed out that relaxing the banking constraint might be considered a competitive alternative to EC4Ps. Such a provision would promote the adoption of new technologies by rewarding early investments. However, a large quantity of banked permits would exacerbate the need for regulator's interventions in response to unexpected shocks of the economy, thus undermining the merits of the  EC4Ps. \vspace{0.20cm}

The first objective of the regulator is to appropriately determine a credible triple $(T,\{N(t)\},P)$ (which need not be unique) so as to curb pollution and to promote firms' incentives to adopt low pollution--emitting technologies in a dynamic fashion. Secondly, in the presence of the European cash--4--permits, the regulator identifies an incentive--equivalent set $\{N(t),P,P_{g},T\}$ such that the policy is credible and, possibly, self--financing (more on this property in Section~\ref{sec:SelfFin}).

\subsection{The firms' characteristics} \label{sec:firmcharact}

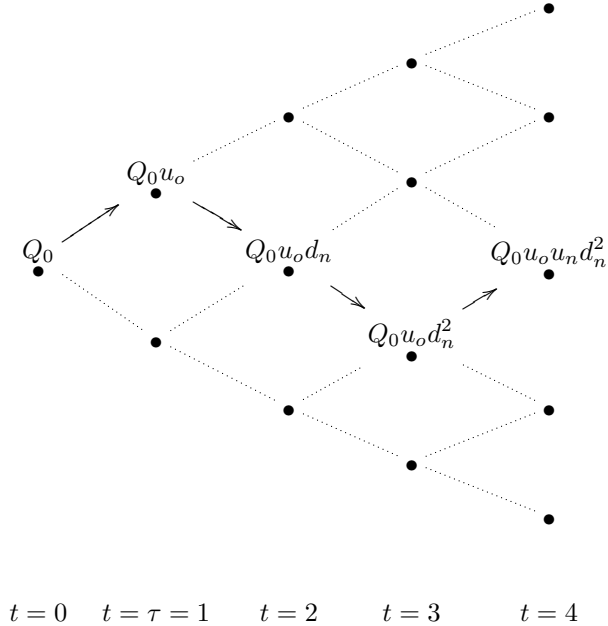
\begin{figure}[htbp]
\begin{center}
$\xymatrix@-1.4pc { & & & & & \stackrel{\displaystyle{}}{
}\\
 & & & & \stackrel{}{\bullet }
 & \\
 & & & \stackrel{}{\bullet }\ar@{.}[ru]
 \ar@{.}[rd] & &\stackrel{\displaystyle{}}{}\\
  & & \stackrel{}{\bullet }\ar@{.}[ru]
  \ar@{.}[dr] & & \stackrel{}{\bullet
  } & \\
   & \stackrel{\displaystyle{Q_{0}u_o}}{\bullet }\ar@{.}[ru]
   \ar@{->}[dr] & &\stackrel{}{\bullet
   }\ar@{.}[ru] \ar@{.}[dr] &
&\stackrel{\displaystyle{}}{
   }\\
   \stackrel{\displaystyle{Q_{0}}}{\bullet }\ar@{->}[ru]
   \ar@{.}[dr] & & \stackrel{\displaystyle{Q_{0}u_od_n}}{\bullet
   }\ar@{.}[ru] \ar@{->}[dr] & &
\stackrel{\displaystyle{Q_{0}u_o u_n d^{2}_n}}{\bullet
   } & \\
    & \stackrel{}{\bullet }\ar@{.}[ru]
    \ar@{.}[dr] & & \stackrel{\displaystyle{Q_{0}u_od_n^{2}}}{\bullet
    }\ar@{->}[ru] \ar@{.}[dr] & &
\stackrel{\displaystyle{}}{
    }\\
     & & \stackrel{}{\bullet }
     \ar@{.}[ru] \ar@{.}[dr] & &
\stackrel{}{\bullet
     } & \\
      & & &\stackrel{}{\bullet
      }\ar@{.}[ru] \ar@{.}[dr] & &
\stackrel{\displaystyle{}}{
      }\\
       & & & &\stackrel{}{\bullet }
        & \\
        & & & & & \stackrel{\displaystyle{}}{ }
        \\
        t=0 & t=\tau=1 & t=2 & t=3 & t=4 & {}}$
\end{center}
\caption{A possible evolution of a firm's cumulative emissions in a four--period long phase, where the firm adopts the new technology at time $t=1.$}
\label{Fig:one}
\end{figure}

Regulated firms are characterized by uncertain emissions and income profiles. The cost of new technology adoption is firm--specific. We assume such investment can only occur once during the regulated phase $[0,T]$, and that it is non--reversible. In order to keep our model tractable, we consider the following binomial dynamics for the (cumulative) emissions $Q^i$ of firm $i$ in unit of pollution:
\be Q^i(t+1)=\left\{
         \begin{array}{ll}
           u^i(t)\cdot Q^i(t), & \hbox{with probability}\,\,q(t). \\
           d^i(t)\cdot Q^i(t), & \hbox{with probability}\,\,1-q(t),
         \end{array}
       \right.
\ee
and $Q^i(0)$ is given. Figure \ref{Fig:one} shows a possible evolution of the cumulative emissions. The factors $u^i(t)$ and $d^i(t)$ denote the production regime of firm $i$ from time $t$ to $t+1;$ since $Q^i$ represents cumulative emissions, we impose $u^i>d^i\ge 1.$ Adoption of low--emitting technology is assumed to affect only the amount of pollution emitted for given output or input, and does not otherwise affect production. The firms' emissions are subject to economic shocks and the implementation of new technologies, among other variables. The former affect a firm's production and are assumed to be exogenous, with the demand for a firm's products contingent on phenomena that are beyond its grasp (a widespread crisis, for example). When demand is high, firm $i$'s cumulative emissions grow by the factor $u^i$, whereas a lower demand is represented by cumulative emissions that increase only by a factor of $d^i.$ On the other hand, the adoption of new technologies is determined endogenously, for example, when potential profits from sales of extra permits or voided penalty costs due to reduced emissions sufficiently compensate investment costs, then firms choose to adopt the new technology. Thus, the incentive to adopt low pollution--emitting technologies depends on the investment costs, the penalty level, the expected supply {and demand for permits} and, implicitly, the expected permit price. As a consequence, in our model both technology adoption and allowance price are generated endogenously and, more importantly, inter--dependently. In particular, the allowance price level at time $t$ depends on the current level of technology adoption, on the current level of uncovered pollution (demand), and the current level of unused allowances (supply). On the other hand, the incentives to adopt new technologies hinge on unanticipated economic shocks and the future values of allowances, the latter being a function of the future permits supply and demand.\vspace{0.20cm}

In order to describe how the adoption of new technologies systematically influences future pollution emissions, we define the stochastic processes:
\be
\begin{array}{cc}
  \mu_o^i(t)= & \left\{
        \begin{array}{ll}
          u_o^i(t), & {\mbox{with probability}}\,q(t) \\
          d_o^i(t), & {\mbox{with probability}}\,1-q(t),
        \end{array}
      \right.
\end{array}
\ee
and
\be
\begin{array}{cc}
  \mu_n^i(t)= & \left\{
        \begin{array}{ll}
          u_n^i(t), & {\mbox{with probability}}\,q(t) \\
          d_n^i(t), & {\mbox{with probability}}\,1-q(t).
        \end{array}
      \right.
\end{array}
\ee
Here $\mu_o^i$ denotes the possible production regimes of firm $i$ under the old technology (hence the subscript ``o'') and $\mu_n^i$ is the analogue under the new  technology. As before, we impose that $\mu_o^i, \mu_n^i\ge 1.$ Under the possibility of technology adoption, the dynamics of the cumulative emissions processes can be expressed as
 \be
 Q^i(t+1)=\mu_{\varsigma}^i Q^i(t),
 \ee
where $\varsigma\in\{u, d\}.$  The probability densities $(q(t), 1-q(t))$ correspond to the likelihood of exogenous shocks that production may experience, then $\mu_t^i$ equals either $\mu_o^i(t)$ or $\mu_n^i(t).$  As described above, one may interpret these shocks as high or low demand phases determined by external factors, which justifies the use of the same probability density for all firms, as well as the fact that the density is independent of the firms' production technologies. We assume that associated with each firm there is a constant cost function $C^i$ defined via
\be
    C^i(\mu_o^i):=0,\quad{\mbox{and}}\quad C^i(\mu_n^i):= C_n^i,
\ee
which indicates the investment cost in which firm $i$ must incur to adopt the new, cleaner technology.\footnote{The adoption of the new technologies generates a {permanent} reduction in pollution emission per unit of input or output. Therefore, it is reasonable to assume the investment cost to be a lump-sum. The case of {a costly choice of different levels of permanent reduction} is left for future research.} Moreover, since production is assumed not to be affected by the adoption of new technology, the firms' aggregate profits from production over $[t, t+1]$ are given by
 \be
 S^i(t+1)=\left\{
         \begin{array}{ll}
           (1+\rho)^{t+1}S_u^i, & \hbox{with probability}\,\,q(t). \\
           (1+\rho)^{t+1}S_d^i, & \hbox{with probability}\,\,1-q(t),
         \end{array}
       \right.
\ee
where $S^i(0)$ is given, $S_u^i, S_d^i>0$ remain constant over the whole regulated phase, and the  appreciation of products is incorporated via the coefficient $(1+\rho)^{t+1},$ with $\rho>r.$  \vspace{0.25cm}

Throughout the remainder of the paper, we shall refer to the quantities
$Q^i(0),$ $\mu_o^i(t),$ $\mu_n^i(t),$  $q(t),$ $C_n^i,$ $S^i(0),$ $S_u^i,$ $S_d^i,$ $r$ and $\rho$ as the {\sl model's primitives}.

\section{Analyzing Firms' Strategic Trading Behavior without Price Support Contracts}

In this section we analyze the impact of the regulator's decisions, i.e.~the triple $\big(T, \{N^i(t)\}_{i\in\I}, P\big)$, on the firms' incentives to adopt the new technologies in the presence of a stand--alone {market for permits.} As mentioned before, the decisions regarding technology adoption take place at the start of each regulated period, whereas the exchange of permits occurs at the end of the period. Each firm's decisions are made taking into consideration its expected permit positions and technology status, both at the end of the period and throughout the remaining duration of the phase. Later we consider the inclusion of specific instruments written on the levels of allowances held at the end of each period.\vspace{0.20cm}

The analysis of the dynamic incentives to adopt new technologies generated by a policy can best be analyzed from the perspective of a regulated firm. However, past literature has concentrated on calculating the aggregate cost savings achieved by regulated firms adopting new technologies. Such an analysis does not showcase an individual firm's incentives to adopt a low pollution--emitting technology. Conversely, in our model the firms' expected (extra) profits and (reduced) losses from participating in a permits system are analyzed given a specific set of parameters $\big(T, \{N^i(t)\}_{i\in\I}, P\big)$. We stress that the regulator's decisions are {made \textit{ex--ante}; in other words, any} information generated throughout the regulated phase provides no feedback to the system's structure. The densities $\{(q(t), 1-q(t)), t \in\{0,\ldots,T-1\}\}$ correspond, therefore, to the a priori beliefs of the regulator. In order to deal with the possible adoption of new technologies, we define the stopping times
\be
\tau^i:=\min\big\{t\in\{0,\ldots,T-1\}\,\mid\,\mu^i(t)=\mu^i_n(t)\big\}.
\ee
To avoid ambiguities we write $Q^i_o(t)$ when production takes place under the old technology, and analogously for $Q^i_n(t).$  Under a permit scheme, a firm is only liable for the amount of non--offset emissions during each period $[t, t+1];$ that is for the difference between the allocated permits and $\Delta_{\varsigma} Q^i(t+1):=Q_{\varsigma}^i(t+1)-Q^i(t),$ where $\varsigma\in\{u, d\}$ specifies whether production over  $[t, t+1]$  takes place in a high or low output regime. Given the dynamics specified above, the cumulative emissions over one period are:
\be
\Delta_{\varsigma} Q^i(t+1)=\left\{
    \begin{array}{ll}
      Q^i_0\big(\prod_{s=0}^{t-1}\mu_o^i(s)\big)\big(\mu_o^i(t)-1\big), & \hbox{if}\,\,t<\tau^i, \\
      Q^i_0\big(\prod_{s=0}^{t-1}\mu_o^i(s)\big)\big(\mu_n^i(t)-1\big), & \hbox{if}\,\,t=\tau^i, \\
      Q^i_0\big(\prod_{s=0}^{\tau^i-1}\mu_o^i(s)\prod_{\tau^i}^{t-1}\mu_n^i(s)\big)\big(\mu_n^i(t)-1\big), &
      \hbox{otherwise}.
    \end{array}
\right.
\ee

\noindent Below we use the notation $h:=(h_1,\ldots, h_m),$ ($h_i\in\{n, o\}$) (the {\sl technology vector}) to indicate the technology under which the firms operate, and $h_{-i}$ stands for ``the technologies of all firms but the $i$-th one''. Hence, at time $t \in \{1,\ldots,T-1\}$ the expected position (in units of allowances) of firm $i$ at time $t+1$ is given by:
\begin{equation}\label{eq:one_period_position}
    \E\Big[\Delta_{\varsigma}Q^i(t+1,h)\Big]-N^i(t),
\end{equation}
where $\Delta_{\varsigma}Q^i(t+1,h)$ represents the {pollution} emissions to be offset at time $t+1,$ contingent on the technology vector being $h$ and the state of the economy being $\varsigma.$

\begin{remark}\label{rmk:states}
Notice first that the technology vector represents a possible combination of the firms' technology adoption strategies. Second, computing the expected cumulative emissions conditioned on the information up to time $t,$ i.e.~$\E\Big[\Delta_{\varsigma}Q^i(t+1)|{\mathcal F}_t\Big],$  is nothing more than a two--terms sum. When this is evaluated at time $t=0$ (as is the case of the regulator) one faces $t+1$
possible states for each combination of the technology vector $h$.
\end{remark}

\subsection{The firms' payoffs} \label{sec:firmspayoff}

As a consequence of the fact that the output quantity is assumed to be unaffected by the technology investment, the technology adoption and the exchange of allowances are the only compliance alternatives of the firms. Let us start by discussing the first {compliance strategy}. At the beginning of each regulated period, each firm that has not adopted a new technology must decide whether it adopts ($n$) or not ($o$). This choice is made evaluating the impact of such investment on the firm's expected payoff stream, considering all possible {\sl technological paths}. Each firm's payoff is clearly affected by the future exchange value of the allowances, which depends on  supply and demand balance of the permit market. To better understand this, let us analyze the different possible permit positions of the firms. In general, a low permit price makes the technology adoption a non--viable strategy. Firms in permit shortage would find compliance by means of allowance purchase a cheaper alternative; firms in permit excess would find it unprofitable to offer their unused permits. Conversely, a high permit price increases potential profits from the sale of extra permits and raises potential compliance costs due to uncovered emissions. In practice, each firm tries to answer the following questions: would the cost of reducing emissions be less than the potential revenues from allowance sales plus avoided penalties? Or would waiting to adopt the new technology and perhaps offset emissions by purchasing cheap allowances be a better strategy? This might provide incentives for some firms to free--ride on the other firms' investments. We stress that decisions about technology adoption are taken under emission uncertainty and in the presence of imperfect competition on the permit market, as described below. \vspace{0.20cm}

The second compliance alternative, the exchange of allowances, is highly dependent on the firms' technology status. The adoption of the new technology reduces emissions and, contingent on the allowance structure $N(t),$ possibly generates excess of unused permits. As observed by \cite{BHQ:95}, a high number of firms that have adopted the low pollution--emitting technologies has a significant effect on the aggregate emissions, thus increasing the potential supply schedule. Quite naturally, in the presence of upward shifts in permit supply, one would expect a low exchange value of the allowances. However, this last statement holds only if sellers automatically offer the entire bulk of their unused allowances. It may be in the self--interest of firms in allowance excess to {limit} their offers, keep the exchange value of permits high and, possibly, collect higher revenues. On the contrary, firms in permit shortage face severe penalties if they fail to deliver an amount of allowances equal to their emissions. Hence, it is in the buyers' best interest to offset all their emissions for any price lower than the penalty level. The firms in permit shortage are, therefore, expected to submit all their demand to the exchange. \vspace{0.20cm}

\begin{remark} Markets for permits are characterized as the exchanges of purely intermediaries titles, where strategic retention of allowances might take place in order to drive the price up. Hence, all what can be transferred, $x^i(t+1, h):=\E\big[\Delta Q^i(t+1, h)\big]-N^i(t)$, is not necessarily allocated. A negative transfer, $x^i(t)<0$, indicates a need for permits. Therefore, the quantities $x^i(t)$ do not necessarily satisfy the traditional market clearing condition $(\sum_{i\in\I}x^i(t) = 0)$ at every period $t\in\{1,\ldots, T\}$. \end{remark}

The discussion above describes two interacting dynamics: the incentives to adopt new technologies and the presence of strategic trading behavior among the firms that are in permit excess. As a consequence of the latter, the analysis of the endogenous technological adoption includes a (non--cooperative) game among firms on the supply--side of the market. The analysis stops either when (or if) all firms operate under the low pollution--emitting technology, or when the regulated phase comes to an end.

\subsubsection{Strategic permits trading and the structure of the allowances' price}

In the present framework, permits are submitted to an exchange and traded exclusively at the end of each period. So far, the only force driving the exchange value of an allowance at the end of each regulated period is the supply and demand balance for permits at that point in time. In principle, the exchange value of the allowance increases as supply (demand) decreases (increases). Below we describe in more detail such dynamics. \vspace{0.20cm}

\begin{figure}
\begin{center}
  \includegraphics[width=7cm]{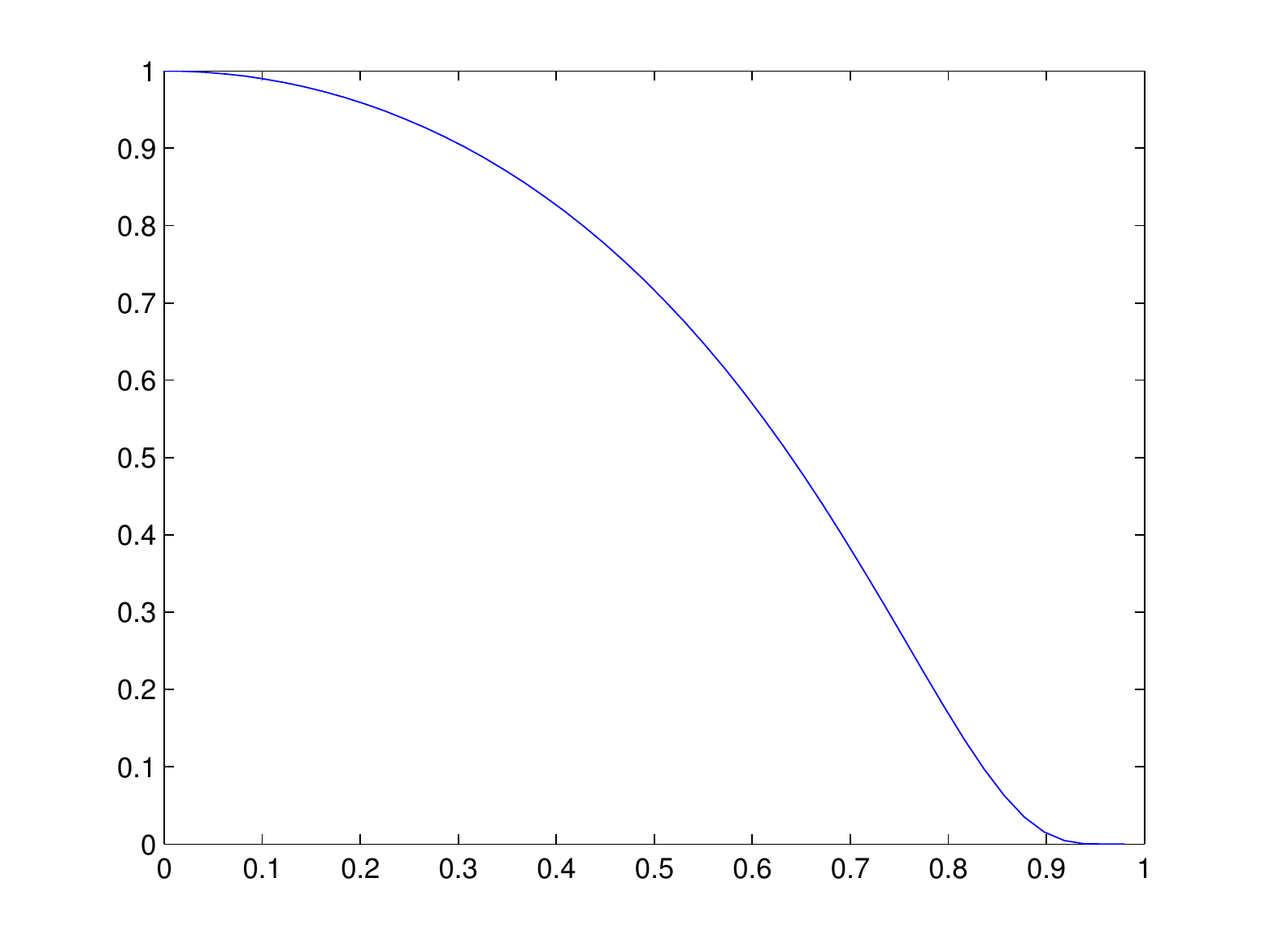}\\
  \caption{The plot of $\eta_1$.}\label{fig:1}
  \end{center}
\end{figure}

We first look at the {\sl{expected}} number of available allowances, for a given technology vector $h.$ The generation of the allowance exchange value at the end of each period will then follow from the knowledge of the state of the world. We start by defining
\be
x^i(t+1, h):=\E\big[\Delta Q^i(t+1, h)\big]-N^i(t).
\ee
This quantity represents the expected position in number of emissions of firm $i$ contingent on the technology vector $h.$ Let
\be 
s(t+1, h):=\Big\{i\in I\,\big|\, x^i(t+1, h)<0\Big\},
\ee
and
\be 
d(t+1, h):=\Big\{i\in I\,\big|\, x^i(t+1, h)\ge 0\Big\},
\ee
be the supply and demand sides of the market (in terms of the firms' expected emissions positions), respectively. The expressions
\be
{\mathcal S}(t+1, h):=-\sum_{i\in s(t+1,
h)}x^i(t+1, h)\quad{\text{and}}\quad {\mathcal D}(t+1,
h):=\sum_{i\in d(t+1, h)}x^i(t+1, h),
\ee
represent the (expected) number of unused permits, i.e.~the aggregate supply, and the (expected) number of non--offset emissions, i.e.~the aggregate demand, respectively. Both expressions are contingent on the technology vector $h.$  We introduce the supply--demand ratio that is later used to determine the allowances' value:
\be {\mathcal R}(t+1,
h):=\left\{
   \begin{array}{ll}
      -\frac{{\mathcal S}(t+1, h)}{{\mathcal D}(t+1, h)}, & \hbox{if}\,\,{\mathcal D}(t+1, h)>0, \\
      0, & \hbox{otherwise}.
   \end{array}
 \right.
\ee
To account for a lower sensitivity of the allowance value in case of extreme permit demand, i.e.~the ratio is close to $0,$ or extreme permit supply, i.e.~the ratio is close to $1,$ we define for $a>0$ the parameterized family of (reaction) functions $\eta_a:[0, a]\to [0,1]$ as
\be \eta_a(x):=\left\{
   \begin{array}{ll}
      \exp\Big\{\frac{x^2}{x^2-a^2}\Big\}, & \hbox{if}\,\,x\in [0, a), \\
      0, & \hbox{otherwise}.
   \end{array}
 \right.
\ee
Figure \ref{fig:1} shows the graph of the function $\eta_a(x)$ for $a=1$.\footnote{The functions $\eta_a$ are the right halves of  scaled mollifiers (see \cite{Evans:08}). They are infinitely smooth at $0$ and $a,$ with derivatives of all orders at these points equal to zero.}  \vspace{0.20cm}

Under a permits scheme, firms in permit shortage face severe penalties, whereas firms in permit excess might profit from the sale of unused permits. By virtue of such a mechanism, it is in the buyers' best interest to offset all their emissions for any price lower than the penalty level. More interestingly, since a lower aggregate supply implies a higher exchange value, it may very well be in the sellers' best interest to reduce the availability of permits. In fact, firms in permit excess must reach a compromise between offering a higher number of cheap permits, or less of them, but at a higher value.\footnote{Notice that this trade--off follows partly from the non--linearity in prices introduced by the functions $\eta_a(x)$.} Such an exchange strategy is part of the sellers' strategic choice. Moreover, each (selling) firm's profit depends not only on its choices, but also on those of the remaining firms on the sell--side. We then face a non--cooperative $m_s$--person game (where $m_s:=\#{\mathcal S}_{\E}(t+1, h))$ when we analyze sellers' decisions on their own supply schedule.\footnote{We assume there is no collusion between firms.} The {exchange} value of the allowances at time $t+1$ (constructed on the base of firms' expected permits positions), given the technology vector $h$ is:
\begin{eqnarray} \nonumber
\Pi(t+1, h)
& := & P\cdot\eta_{{\mathcal R}(t+1, h)}\Big(-\frac{^e{\mathcal S}(t+1, h)}{{\mathcal D}(t+1, h)}\Big) \\
& = & P\cdot\exp\Big\{\frac{^e{\mathcal S}(t+1, h)^2}{^e{\mathcal S}(t+1, h)^2-{\mathcal D}(t+1, h)^2}\Big\}.
\end{eqnarray}
where the quantity $^e{\mathcal S}(t+1, h)$ represents the total number of unused permits submitted to the exchange and available for sale.

\begin{remark} Since we assume that the penalty is an alternative to compliance, this quantity represents an upper bound for the price of an allowance. In particular, one should observe that the indifference buy--price for an allowance that safeguards a firm that is in permits shortage from paying the penalty $P$ is precisely $P$. Hence, by construction $0\le\Pi(t+1, h)\le P.$ \end{remark}

\begin{figure}[hb!]
\begin{center}
\includegraphics[width=10cm]{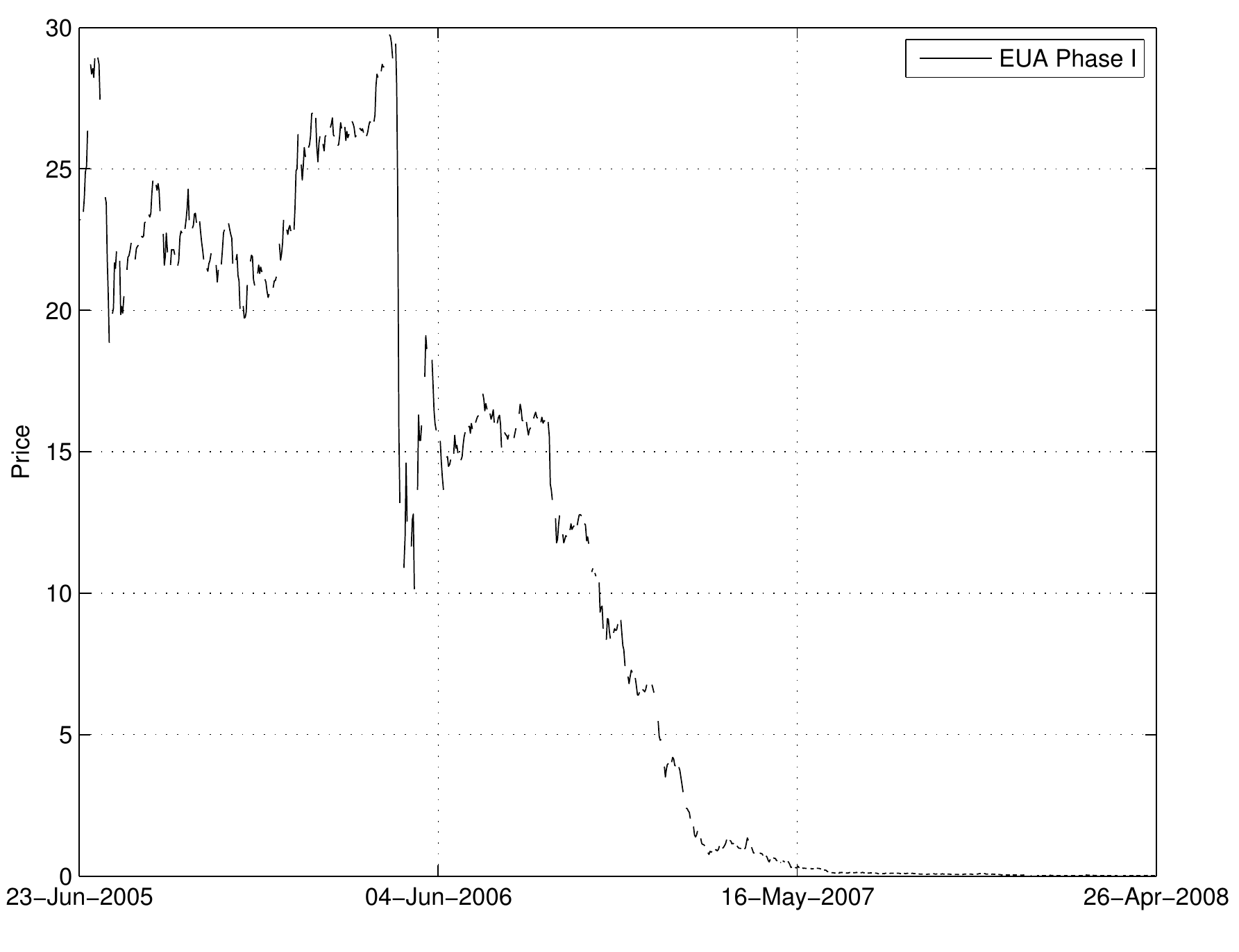}\\
\caption{Spot price of the EU Allowance Unit from 2005 until 2008 on the European Climate Exchange (ECX).}\label{fig:3}
\end{center}
\end{figure}

\noindent Before we proceed with our analysis, we shortly analyze an empirical example.
\begin{example}
Let us consider the evolution of the allowance price in the first regulated phase of the European Union Emission Trading Scheme. Figure \ref{fig:3} represents the evolution of the allowance price on an exchange market during Phase I. By the end of April 2006, it had become apparent that the number of permits required to offset the expected emission, $\E\Big[\sum_{i\in\I}\sum_{t=0}^TQ^i(t)\Big],$ where $T$ corresponds to beginning of 2008, was largely overestimated. In other words, the total number of allocated permits, $N = \sum_{t=2005}^{2008}N(t),$ was largely sufficient. Hence, the large downwards jump in the proximity of April 2006. However, the allowance exchange value remained for some time far away from zero. Arguably, this reflects the unwillingness of firms in permit excess to offload their surplus at low valuation levels. By refraining from offering {the entire amount of} extra permits, firms in permits need faced a relatively high sustained allowance price for quite some time.
\end{example}

\subsubsection{Determining the aggregate supply $^e{\mathcal S}(t+1, h)$ and the permit price $\Pi(t+1, h)$}\label{subsect:prices-no-cash}

In order to analyze how sellers choose their supply schedules, we consider the income generated from the permit exchange of firm $i\in s(t+1, h)$ as a function of the {\it supply vector} $\big({^e}x^i(t+1, h), {^e}x^{-i}(t+1, h)\big),$ i.e.~the number of allowances the $i$--th firm would submit to the exchange, and those that would be submitted by the other firms on the sell--side. Namely
\begin{equation}\label{eq:gains-per-firm}
\Psi^i\big({^e}x^i, {^e}x^{-i}\big) := {^e}x^i P\cdot\eta_{\mathcal R}\Big(-\frac{^e{\mathcal S}^{-i}+{^e}x^i}{{\mathcal D}}\Big),
\end{equation}
where $^e{\mathcal S}^{-i}=\sum_{j\in s\setminus\{i\}}{^e}x^j,$ and we have omitted the arguments $(t+1, h)$ to keep the notation as uncluttered as possible. Equation~\eqref{eq:gains-per-firm} represents the trade--off described above: sellers have to choose between offering a higher number of cheap permits, or less of them, but at a higher value. Once all sellers' offers are collected, the {\it low--side} or {\it high--side} of the market is determined. The terms low-- and high--side of the market denote whether supply (or demand) exceeds or not demand (or supply), respectively. It is in the interest of sellers that the supply remains the high--side of the market. \vspace{0.20cm}

Below, we follow the well--worn trail of studying the best--response correspondences of each seller's response to the remaining sellers' submission of allowances to the exchange, and we show the existence of Nash equilibria of this strategic interaction. To this end we have the following:
\begin{lemma}\label{lm:single-valued} For any supply vector $\big({^e}x^i, {^e}x^{-i}\big),$ the mapping $
{^e}\tilde{x}^i\mapsto\Psi^i\big({^e}\tilde{x}^i, {^e}x^{-i}\big)$
is maximized at a single point. In other words, the correspondence
\be
\Phi^i({^e}x^i, {^e}x^{-i})={\text argmax}\Big\{\Psi^i\big({^e}\tilde{x}^i, {^e}x^{-i}\big)\,\big|\,{^e}\tilde{x}^i\in [0, x^i]\Big\}
\ee
is single valued.
\end{lemma}
\begin{Proof} We assume that ${\mathcal D} >0,$ otherwise there is no  demand for allowances, hence no market, and the maximizer is trivial. We must show that the mapping
\be
x\mapsto x\exp\left\{\Big(\frac{K_1+x}{K_2}\Big)^2\Big/\Big(\Big(\frac{K_1+x}{K_2}\Big)^2-b\Big)\right\},
\ee
where $b=\Big(\frac{K_1+ x^i}{K_2}\Big)^2,$ $K_1=^e{\mathcal S}^{-i}$ and $K_2={\mathcal D}$ is maximized at a single point of $[0, {x}^i].$  By rescaling if necessary, we may assume without loss of generality that $K_2=1.$ Moreover, we may assume that $K_1+x^{i}\le 1,$ given that under the previous assumption $\eta_{{\mathcal R}}\equiv 0$ for any value larger than $1.$ Initially we assume that $K_1+x^{i} = 1,$ i.e.~firm $i$ has the ability, given $K_1,$ to fully satisfy the demand for allowances. Since in such a case the values of the mapping under investigation are strictly positive on $(0, {x}^i)$, we need only to seek interior maximizers. The first order conditions yield the equation
\be
L(x):=(K_1+x)^4-4(K_1+x)^2+2K_1(K_1+x)+1=0.
\ee
We have that $L(0)=(K_1-1)^2>0,$ and $L(x^i)=-2+2{K_1}<0.$ It follows from the Intermediate Value Theorem that $L$ has a root $x^i_0$ in $(0, x^i).$ To show uniqueness, we note that $L''$ changes sign only once on $[0, \infty),$ which given the general shape of the graph of a fourth--degree polynomial implies there are only two roots in this interval. Since $L(x^i)<0$ and $\lim_{t\to\infty} L(t)=\infty,$ we conclude that the remaining root lies beyond $x=x^i.$ If it were the case that $K_1+x^i<1,$ then either $x^i_0\le K_1+x^i,$ in which case the previous result holds, or the maximizer is precisely $x^i.$
\end{Proof}

\noindent Notice that of the three requirements to apply Kakutani's Fixed--point Theorem (see for example~\cite{Meyerson:91}), Lemma~\ref{lm:single-valued} takes care of the non--vacuity and the convexity. We then need an upper--semicontinuity result, which we present in the following.

\begin{lemma}\label{lm:usc} Let the mapping $\Phi:\re^{m_s}\to\re^{m_s}$ be defined via
\be
\Phi(x^{i_1},\ldots, x^{i_{m_s}}):=\bigotimes \Phi^i(x^{i_j}, x^{-i_{j}})
\ee
for $(x^{i_1},\ldots, x^{i_{m_s}})\in\bigotimes [0, x^{i_j}],$ then $\Phi$ is continuous.
\end{lemma}
\begin{Proof} For $x\in [0, x^i],$ the mapping
\be
K_1\mapsto{x}\exp\left\{\Big(\frac{K_1+x}{K_2}\Big)^2\Big/\Big(\Big(\frac{K_1+x}{K_2}\Big)^2-b\Big)\right\}
\ee
is continuous. Notice that $x^{-i_{j}}$ is a relevant statistic for $\Phi^i(x^{i_j}, x^{-i_{j}})$ only  through
$\sum_{k\neq j} x^{i_{k}},$ and clearly the mapping
\be
(x^{i_1},\ldots, x^{i_{m_s}})\mapsto\bigotimes\sum_{k\neq j} x^{i_{k}}
\ee
is continuous. It follows immediately that the mapping
$
(x^{i_j}, x^{-i_{j}})\mapsto\Phi^i(x^{i_j}, x^{-i_{j}})$
is continuous over $\bigotimes [0, x^{i_j}],$ which finalizes the proof.

\end{Proof}

\noindent Lemmas~\ref{lm:single-valued} and~\ref{lm:usc}, together with Kakutani's Fixed--point Theorem imply that the mapping
\be
(x_1,\ldots,x_{m_s})\mapsto\Phi(x_1,\ldots,x_{m_s})
\ee
has a fixed point. In other words, we have proved the following

\begin{thm}\label{thm:nash} The (non--cooperative) game ${\mathcal G}=\big\{[0, x^i_h], \Psi^i\big\}_{i\in {\mathcal S}}$ possesses a pure--strategy Nash Equilibrium.
\end{thm}

\noindent Moreover, the Nash equilibrium mentioned in Theorem~\ref{thm:nash} is unique. This follows from the fact that it {coincides with} the unique solution of the system of equations:
\be
D_{y^{i_j}}\Psi(y^{i_j}, y^{-i_j})+\lambda^{i_j}=0,\quad \lambda^{i_j}(y^{i_j}-x^{i_j})=0,\quad\sum_{j}x^{i_j}\le 1,\quad j=1,\ldots, m_s,
\ee
where the $\lambda^{i_j}$'s are the Lagrange multipliers associated to the constraints $y^{i_j}-x^{i_j}\le 0.$ From now on $^{*}x^{i_j}(t+1, h)$ will represent the $j$--th entry of the Nash--equilibrium that results from the solution of the game among the firms in permit excess, contingent on the technology vector $h.$
The {\it expected equilibrium exchange value} of an allowance, contingent on the technology vector $h$ is:
\be
\Pi^*(t+1, h):=P\cdot\eta_{{\mathcal
R}(t+1, h)}\Big(-\frac{^*{\mathcal S}(t+1, h)}{{\mathcal D}(t+1, h)}\Big).
\ee

A further comment should be made regarding the generation of prices in the case where some of the constraints $y^{i_j}-x^{i_j} \le 0$ prove to be binding. If $x^{i_j}=$ $^{*}x^{i_j},$ then the following point is of course moot. Otherwise, if some firms are not able to increase their supply schedules up to the (unconstrained) equilibrium level, then the firms that still have availability of permits have extra room to increase their exchange offers. Interestingly, the additional potential number of permits does not restore the original aggregate supply of the unconstrained problem, i.e.~the aggregate supply of permits (in equilibrium) in the presence of binding constraints is bound above by that of the unconstrained problem. When some firms' supply--constraints bind, therefore, the equilibrium price increases. Moreover, the firms with non--satiated constraints collect higher profits than in the unconstrained case by virtue of a lower (aggregate) supply. We formalize these claims in the following

\begin{lemma}\label{lm:Higher-price} {Let $\{^{*}x^{i_j}(t+1, h)\}$ be the equilibrium supply profile  of the game ${\mathcal G}=\big\{[0, 1], \Psi^i\big\}_{i\in {\mathcal S}_{\E}},$ then the equilibrium supply profile $\{\tilde{x}^{i_j}(t+1, h)\}$ of the constrained game ${\tilde{\mathcal G}}=\big\{[0, x^i_h], \Psi^i\big\}_{i\in {\mathcal S}_{\E}}$ (with $x^i_h<1$), and the corresponding price $\tilde{\Pi}(t+1, h)$ satisfy:}

\begin{enumerate}

\item If $x^i_h <$  $^{*}x^{i}(t+1, h),$ then $\tilde{x}^{i}(t+1, h) = x^i_h.$

\item If $x^i_h >$ $^{*}x^{i}(t+1, h),$ then $\tilde{x}^{i}(t+1, h)>$$ ^{*}x^{i}(t+1, h).$

\item $\tilde{\Pi}(t+1, h) > \Pi^*(t+1, h).$

\end{enumerate}

\end{lemma}
{\begin{Proof} The first point follows from the fact that the best--response path of a firm whose supply--constraint is binding reaches and is absorbed by  the corresponding $x^i_h.$ Next we assume $\#s=2$ for notational simplicity. Since for a fixed $x_1,$ the expression
\be
1-\frac{2x_1(x_1+x_2)}{((x_1+x_2)^2-1)^2},
\ee
which corresponds to the first order conditions of firm $1,$ is decreasing in $x_1,$ if $x^1_h < ^{*}x^{1}(t+1, h),$ then $\tilde{x}^{2}(t+1, h)>\, ^{*}x^{2}(t+1, h).$ In what follows we drop the arguments $(t+1, h)$ for clarity. The question remains whether or not the increased supply by firm $2$ over the unconstrained--equilibrium level fully compensates the decreased supply of firm $1,$ as to leave aggregate supply unchanged.  The answer is no. If firm $2$ were to offer $^{*}x^{1}-\tilde{x}^{1}+\,^{*}x^{2},$ we would have
\be
1-\frac{2\big(^{*}x^{1} - \tilde{x}^{1} +\,^{*}x^{2}\big)\big( ^{*}x^{1} +\,^{*}x^{2} \big)}{\big(\big( ^{*}x^{1} +\,^{*}x^{2} \big)^2-1\big)^2} = 1-\frac{2\,  ^{*}x^{2} \big( ^{*}x^{1} +\,^{*}x^{2} \big)}{\big(\big( ^{*}x^{1} +\,^{*}x^{2} \big)^2-1\big)^2}-\frac{\big(^{*}x^{1} - \tilde{x}^{1}\big)\big( ^{*}x^{1} +\,^{*}x^{2} \big)}{\big(\big( ^{*}x^{1} +\,^{*}x^{2} \big)^2-1\big)^2} < 0.
\ee
The inequality follows from the fact that the first two terms on its left hand side add up to zero (the first order condition for the unconstrained equilibrium) and $^{*}x^{1} - \tilde{x}^{1}>0.$ We conclude that $\tilde{x}^{1}+\tilde{x}^{2} < ^{*}x^{1} +\,^{*}x^{2},$ which in turn implies $\tilde{\Pi}(t+1, h) > \Pi^*(t+1, h).$
\end{Proof}}

\subsubsection{The mechanics of exchange order execution}\label{sscec:mech}

In this section we study how exchanges of allowances are executed in the {permits} market. We need, therefore, to specify how buy-- and sell--orders are matched. Offers are submitted to an exchange and allowances are traded exclusively at the end of each period, hence the state of the economy is known at the time of trading. Notation is carried on from the previous section, but we refer now to realized positions. All orders are submitted to a centralized exchange market, in which they are randomly (and uniformly) matched one--by--one. Since it is in the interest of sellers that the supply side remains the high--side of the market and they can strategically keep it that way, we can expect that the buy side of the market will be the low one. As a consequence, all of the sellers' orders get executed. Furthermore, the probability that the orders of firm $i\in d(t+1, h)$ are matched is
\be
-\dfrac{x^i(t+1, h)}{{\mathcal D}(t+1, h)}.
\ee
In other words, firm $i$'s access to the sell--side of the market corresponds to its relative contribution to the aggregate demand schedule. In view of the latter, the executed orders of firm $i$ are:
\be
X^i(t+1, h)=\dfrac{x^i(t+1, h)}{{\mathcal D}(t+1,h)}\cdot ^{e}{\mathcal S}(t+1, h).
\ee

\subsubsection{The firms' expected payoffs over $[t, t+1]$ from permit exchange.}

Recalling the discussion at the beginning of Section~\ref{sec:firmspayoff}, the decisions regarding the adoption of new technologies and trading permits have inter--connected dynamics. In fact, although allowances are traded on the exchange market at the end of each regulated period, it is the firms' expected payoffs that play the crucial role of determining the $\tau^i$s. Hence, it is the firms' expected payoffs that drives the {dynamic} incentives to adopt the low pollution--emitting technology. In this section we analyze the firms' payoffs over a single period. This will be the building block for our study of the firms behavior regarding the {dynamic} adoption of the new technology. We assume that the firms' expectations on how their orders will be matched correspond to those presented in Section~\ref{sscec:mech}. \vspace{0.20cm}

Let $\phi^i(t+1, h)$ denote the (expected) payoff for firm $i,$ contingent on the technology vector being $h.$  If $i\in s(t+1, h),$ then firm $i$ would expect to sell $|^{*}x^i(t+1, h)|$ allowances, and its profit would be
\be
\phi^i(t+1, h)=\Pi^*(t+1, h)\cdot|^{*}x^i(t+1, h)|+\Delta S^i(t+1).
\ee
On the other hand, if $i\in d(t+1, h),$ then
\be
X^i(t+1, h)=\dfrac{x^i(t+1, h)}{{\mathcal D}(t+1, h)}\cdot ^*{\mathcal S}(t+1, h),
\ee
and the firm's expected profit would be
\be
\phi^i(t+1, h)=\Delta S^i(t+1)-P\cdot x^i(t+1, h)\dfrac{{\mathcal D}(t+1, h) -\, ^{*}{\mathcal S}(t+1, h)}{{\mathcal D}(t+1, h)}-\Pi^*(t+1, h)\cdot X^i(t+1, h).
\ee

\noindent The quantity $x^i(t+1, h)\big({\mathcal D}(t+1, h)-{\mathcal S}(t+1, h)\big)/{\mathcal D}(t+1, h)$ represents the number of emissions that are not offset by allowances, and for which the the prescribed penalty would be levied.

\subsubsection{The firms' expected payoffs over $[t_0, T].$}\label{subsubsec:expectedProfitsFinal}

In this section we describe the mechanism that governs the firms' investment decisions. We stress {once again} that the incentive to adopt the new technology depends on the firm's potential profits and avoided penalty costs. In order to quantify this amount, {each} firm computes its corresponding expected payoff for the remainder of the regulated phase, which shall be denoted by $[t_0, T],$ over all possible technology vector scenarios (see Remark~\ref{rmk:states}). Let us first introduce the following definition:
\be
\O(t_0):=\big\{i\in\I\,\mid\, \mu^i(t_0-1)=\mu_o^i(t_0-1) \big\}.
\ee
where $\O(t_0)$ represents the set of firms that have not invested in the new technology up to time $t_0-1.$ When $i\in\O(t_0),$ firm $i$ must make the choice at time $t=t_0$ to invest or wait. A family of firm--specific and concave utility functions is used to assess if the adoption of new technology at time $t=t_0$ is economically viable or not:
\be
\Upsilon^i:\{0,\ldots, T-1\}\times\{n, o\}\to\re.
\ee
These are constructed within the von Neumann--Morgenstern expected utility paradigm using as basis the concave functions
\be
U^i:\re\to\re.
\ee
\begin{figure}
\be
\begin{array}{cc}
 {\text {this firm adopts at time}}\;t_0  \longrightarrow & \left(
\def\objectstyle{\scriptstyle}
\def\labelstyle{\scriptstyle}
\vcenter{\xymatrix @-1.2pc {
0  & 1 & 0 & \cdots & 0  \\
1  & 0 & \cdots & \cdots & 0\\
0  & 0 & 0 & \cdots & 1}}\right)
\end{array}
\ee\vspace{.1in}
\caption{A possible representation of a path matrix with three remaining firms, $\#\O(t_0)=3,$ and where the second firm adopts the new technology at time $t_0$. We denote such a matrix $\mm^2_n(t_0)$.}\label{fig:pathmatrix}
\end{figure}

In order to model all the possible scenarios over the $[t_0, T]$ period, we consider the matrices of dimension $\#\O(t_0)\times(T-t_0),$ where each row contains a single $1$ and the rest of its entries are $0$'s.\footnote{Since a $1$ in the $(T-t_0)$--th column denotes that the corresponding firm did not adopt the new technology before the end of the phase, there is no loss of generality in assuming the matrices are row--stochastic.} We shall call such matrices  {\sl path matrices}. Each possible matrix with this structure denotes a possible way in which technology adoption may be undertaken by the $\#\O(t_0)$ firms that can still make such decision. A $1$ in the $(j,t)$--th entry represents technology adoption of firm $j$ at time $t.$ Figure \ref{fig:pathmatrix} represents one of such possible matrices. We make the distinction between those matrices whose $(i, 0)$--th entry is $1$ (firm $i$ adopts the new technology at time $t_0$) and those whose $(i, 0)$--th entry is $0$ (firm $i$ has decided to wait), and we denote these sets $\mm^i_n(t_0)$ and $\mm^i_o(t_0)$ respectively.  In terms of cardinality, $\#\mm^i_n(t_0)=(T-t_0)^{\#\O(t_0)-1}.$ This is not the case for $\mm^i_o(t_0).$  The cardinality of this set is $(T-t_0)^{\#\O(t_0)}-(T-t_0)^{\#\O(t_0)-1}.$ We recall that if the $(j, T-t_0)$--th entry of a path matrix is $1,$ then the matrix represents a scenario where the $j$--th firm would not adopt low--emitting technology before the end of the phase. \vspace{.2in}

Let $M\in\mm^i_n(t_0),$ in order to compute firm $i$'s payoff (should this matrix represent the way in which adoptions are undertaken), we construct a sequence of technology vectors $\big\{h_M(t)\big\}_{t=t_0}^{T-1}$ by defining their entries via:
\be h_M(t)_j=\left\{
                \begin{array}{ll}
                  n, & \hbox{if}\,\,M(j, t)=1, \\
                  o, & \hbox{if}\,\,M(j, t)=0\;\hbox{or}\;j\notin\O(t_0).
                \end{array}
              \right.
\ee
Notice that if $M\in\mm^i_n(t_0),$ then $h_M(t_0)_i=1.$

\begin{remark}
Notice first that, conditional on the information generated until time $t=t_0,$ the emissions levels $Q^i(t_0)$ are deterministic. However, the quantities $Q^i(t)$ ($t>t_0$), which are required to compute the {firms'} future payoffs, are random. Hence, a path matrix $M$ describes a possible evolution of the technological vector given the past decisions, and therefore a possible evolution of the $x^i$'s. This in turn implies that, a priori, each choice of $M$ determines a particular stream of expected prices generated by the firms' expected positions in terms of emissions. Only at the end of each period the quantities ``realized emissions minus holdings in permits'' generate a unique exchange price for the permits. Since information generated throughout provides no feedback to the {dynamic incentives structure of the} permits system, all of regulator's decisions are based on the streams $\{\Pi^*(t, h), t\in[1,T]\}.$
\end{remark}

For the evaluation of firms' incentives to adopt the new technologies, we use the payoffs expected values, corresponding to the possible $h_M(t)$'s. The payoff we are seeking is then:
\begin{equation}\label{eq:change}
{\mathcal P}^i_M(t)=\sum_{t=t_0}^{T-1}
\phi^i(h_M(t), t)
-(1+r)^{T-t_0}C^i.
\end{equation}
The first term of the sum above represents the per--period payoff stream associated to the path matrix $M;$ whereas $(1+r)^{T-t_0}C^i$ is the cost of change, taken to time $t=T.$ When $M\in\mm^i_o(t_0)$ the situation is quite similar, except the possible adoption of the new technology at some future period. This is the time $t=\tau^i,$ which was defined previously. The payoff associated to this $M$ is:

\begin{equation}\label{eq:nochange}
{\mathcal P}^i_M(t)=\sum_{t=t_0}^{\tau^i-1}\phi^i(h_M(t), t)+\sum_{t=\tau^i}^{T-1} \phi^i(h_M(t), t) -(1+r)^{T-\tau^i}C^i.
\end{equation}
The only difference between expressions \eqref{eq:change} and \eqref{eq:nochange} is the fact that in the latter $\tau>t_0$ and accordingly the future cost of change kicks in after $t=t_0.$ For $k\in\{o, n\},$ we define the {\sl payoffs vector} associated to $\mm^i_k(t_0)$ as the vector with entries ${\mathcal P}^i_M(t)$ ($M\in\mm^i_k(t_0)$) ordered in increasing fashion, and we denote the latter by ${\mathcal V}^i_k(t_0).$ Firm $i$ then assigns the following rating to the technological option $k:$
\be
\Upsilon^i(t_0, k):=\big(1/\#\mm^i_k(t_0)\big)\sum_{j=1}^{\#\mm^i_k(t_0)}U^i\big({\mathcal V}^i_k(t_0)_j\big).
\ee
If $\Upsilon^i(t_0, n)\ge\Upsilon^i(t_0, o),$ then firm $i$ adopts the low--emitting technology at time $t_0,$ otherwise it waits.

\begin{remark}
The concavity of $U^i$ represents firm $i$'s pessimism, and  $\Upsilon^i(t_0, k)$ is its expected utility under the assumption that all states of the world (represented by the elements of  $\mm^i_k(t_0)$) are equiprobable.
\end{remark}

We postpone the presentation of some examples until Section~\ref{ssec:ExEC4P}. There we first discuss the permit exchange dynamics. Then, we compare the evolution of the technological vectors -timing and level of technology adoption-  within the framework presented above, and under the presence of a price support instrument (the European Cash--4--Permits) that we introduce below.

\section{Promoting  Dynamic Technology Adoption}\label{sec:EC4P}

Since the allowance schedule is set \textit{ex--ante}, the deterioration of the economy or its improvement may provide incentives for the regulator to adjust the level of the policy. This would clearly undermine the credibility of the policy. {Following \cite{LT:96b}, we implement a price--support instrument.} More precisely, we introduce a free--of--charge option contract that we call {\it European--Cash--4--Permits} (EC4P for shorthand). The EC4P is a put--type option contract written on the final holdings of permits and contingent on the technology status. {\cite{BHQ:95} and \cite{KennedyLaplante:99} have proposed a similar type of policy, where the regulator has the ability to buy back permits. The rationale behind the introduction of such a policy instrument is to sustain the credibility of the policy by reducing the need for the regulator's intervention, and ultimately restore the dynamic incentive to adopt low pollution--emitting technologies. It should be noted that under this policy the outstanding number of permits is modified via actions of the firms, and not due to direct intervention by the regulator. One may think of the EC4Ps as a (floating)} minimum price guarantee contingent on the technology status. At maturity, this instrument guarantees a per--permit amount of money, $P_g$, if and only if the firm ends in permit excess and contingent on adoption of new technology. \vspace{0.20cm}

We assume that the adoption of low--emitting technologies is perfectly verifiable by the regulator, thus ruling out moral hazard. All firms that have adopted the new technology have access to EC4Ps. In fact, their investment entitles these firms to as many EC4P options as allowances they hold. These options, however, are non--transferable, and they can only be exercised within the regulated period they are issued. The regulator establishes the minimum price guarantee, $P_{g},$ that a firm operating under the new technology receives per each allowance returned together with an EC4P contract. The level $P_{g},$ which clearly ranges between zero and the penalty $P,$ is a new policy variable under the regulator's control. Below we show that the presence of EC4Ps creates a (floating) price floor for the exchanged permits. The rationale behind this price floor is the fact that the indifference sell--price for an allowance offered by a firm that adopted the new technology, and which is in permit excess is precisely $P_{g}$. This mechanism may restore the incentive to adopt new technology.

\subsection{The impact of the EC4P on the aggregate supply and the permit price}

The introduction of the EC4P has an impact on the number of the permits that firms in permit excess will submit to the exchange and, therefore, affects the allowance exchange value. We maintain the notation $x^i(t+1, h),$  ${\mathcal D}(t+1, h)$ and ${\mathcal S}(t+1, h)$ used in the previous sections. If firm $i$ has adopted low pollution--emitting technology and it is in permit excess, then the quantity $x^i(t+1, h)$ can be divided into $^{e}x^i(t+1, h)$ and $^{c}x^i(t+1, h).$ The former indicates the number of permits submitted to the exchange, and the latter those that are ``cashed--4--permits''. \vspace{0.20cm}

In parallel to Section~\ref{subsect:prices-no-cash}, we must now study the generation of allowance prices considering how firms in permit excess balance their positions in EC4P and market--exchanged permits. Assume that firm $i$ is in permit excess and that it operates under new technology. Then given that the other firms in excess have submitted $^e{\mathcal Ts}^{-i}=\sum_{j\neq i}{^e}x^{j}$ permits into the exchange market, firm $i$'s choice of ${^e}x^i$ yields a profit
\begin{equation}\label{eq:gains-per-firm-EC4P}
^{4}\Psi^i\big({^e}x^i, {^e}x^{-i}\big) := P_{g}(x^i-{^e}x^i)+ {^e}x^i P\cdot\eta_{{\mathcal
R}}\Big(-\frac{^e{\mathcal S}^{-i}+{^e}x^i}{{\mathcal D}}\Big).
\end{equation}

Notice that the mapping  $x\mapsto P_{g}(x^i-x)$ has constant slope $-P_{g}.$  A too high choice of $P_{g}$ could then result in ${^e}x^i\equiv 0$ being the optimal exchange strategy for all firms that are in permit excess, and which operate under the new technology. It is clear that the condition $P_{g}<P$ should hold, otherwise the regulator would offer arbitrage opportunities to some regulated firms. In fact this condition is sufficient to guarantee that markets will not shut down, in the sense that it will not be optimal for all the firms to submit zero--supply schedules and exercise their EC4P. The latter claim follows from the fact that
\be
\frac{d}{dx}\, ^{4}\Psi^i\big(x, 0\big)\Big|_{\{x=0\}}=P-P_{g}.
\ee
If all other firms were to exercise their EC4Ps, the marginal utility of firm $i$ at zero would be increasing in its submissions in the exchange, and it would find it suboptimal to abstain from trading permits. \vspace{0.20cm}

For the ease of exposure, let us now split the sets $s(t+1, h)$ into $s^o(t+1, h)$ and $s^n(t+1, h).$ These sets represent, respectively, the firms that would be (contingent on $h$) in permit excess at the end of the period $[t, t+1]$ and that would operate under the old technology throughout the period, and those that would be in permit excess, but which had already adopted the new technologies. Firms that belong to $s^o(t+1, h)$ simply operate as before; however, those in $s^n(t+1, h)$ will not submit permits into the exchange unless they can make at least $P_g$ per unit of allowances traded. We again assume without loss of generality that total demand equals one, so that the payoff of a firm in $s^n(t+1, h),$ which submits ${^e}x^i$ to the exchange, given that the other firms in permit excess have submitted $K_1$ is:
\be
^{4}\Psi^i\big({^e}x^i, {^e}x^{-i}\big) = P_{g}(x^i-{^e}x^i)+ {^e}x^i P\cdot\exp\left\{\frac{(K_1+{^e}x^i)^2}{{(K_1+{^e}x^i)^2-1}}\right\}.
\ee
Similarly to Lemma~\ref{lm:single-valued}, we have the following
\begin{lemma}\label{lm:single_val_EC4P}
For $i\in s^n(t+1, h),$ and any supply vector ${^e}x^{-i},$ the mapping $x\mapsto ^{4}\Psi^i\big(x, {^e}x^{-i}\big)$ is maximized at a single point of $[0, x^i).$
\end{lemma}
\begin{Proof} As in the proof of Lemma~\ref{lm:single-valued}, we first assume that $K_1+x^i = 1.$ Let
\be
f(x) := P_{g}(x^i-x)+ x P\cdot\exp\left\{\frac{(K_1+x)^2}{{(K_1+x)^2-1}}\right\}
\ee
and
\be
g(x) := f'(0) x + P_g x^i.
\ee
The graph of the function $g$ in simply the tangent to the graph of $f$ at $x = 0.$ We observe that for $x\in (0, x^i),$
\begin{eqnarray*}
g(x) - f(x) & = & (f'(0)+ P_g)x- x P\cdot\exp\left\{\frac{(K_1+x)^2}{{(K_1+x)^2-1}}\right\}\\
            & = & x P\cdot\exp\left\{\frac{(K_1)^2}{{(K_1)^2-1}}\right\}-x P\cdot\exp\left\{\frac{(K_1+x)^2}{{(K_1+x)^2-1}}\right\}\\
            & > & 0.
\end{eqnarray*}
In other words, the graph of $f$ is strictly under the graph of its tangent at $x = 0.$
We also have that $f'(0)= -P_g+ P\cdot\exp\left\{\frac{(K_1)^2}{{(K_1)^2-1}}\right\}.$ If this quantity were to be non--positive, then $f$ would be maximized at $x = 0.$ On the other hand, if $f'(0)>0$ and $f'(x^i)<0,$ then there is $x_0^i\in (0, x^i)$ such that $f'(x_0^i)=0.$ Moreover, $x\mapsto x P\cdot\exp\left\{\frac{(K_1+x)^2}{{(K_1+x)^2-1}}\right\}$ is a quasiconcave (thus single--cusped) mapping, hence so is $x\mapsto f(x).$ We may then conclude that $x_0^i$ is unique. The case where $K_1+x^i<1$ follows in a similar fashion, but $f(x^i)>0.$ Nevertheless, $f$ remains quasiconcave, hence maximized at a single point.

\end{Proof}

With Lemma~\ref{lm:single_val_EC4P} at hand, the analysis of the existence of equilibria of the game  ${\mathcal G}=\big\{[0, x^i_h],\, ^{4}\Psi^i\big\}_{i\in {\mathcal S}}$ is analogous to that of Section~\ref{subsect:prices-no-cash}, and the corresponding equilibrium will be denoted by $\{^{4}x^i\}_{i\in {\mathcal S}}.$ We may then conclude the existence of a (unique) equilibrium price $^{4}\Pi^*(t+1, h),$ which in turn keeps our description of the mechanics of trading mostly unchanged. The notable difference being that a firm in permit excess operating under the new technology, has a profit
\be
^{4}\phi^i(t+1, h)=\,^{4}\Pi^*(t+1, h,)\cdot|^{4}x^i(t+1, h)|+P_g|x^i(t+1, h)-\,^{4}x^i(t+1, h)| +\Delta S^i(t+1).
\ee
contingent on the technology vector $h.$ Furthermore, {as shown in Figure \ref{figure:3},} we have the following

\begin{proposition}\label{prop:greaterProfits}
Under identical primitives and identical triples $\big(T, \{N^i(t)\}_{i\in\I}, P\big),$ the following holds for all $t\in [0, T]:$
\be
^{4}\Pi^*(t+1, h) \ge \Pi^*(t+1, h).
\ee
\end{proposition}
\begin{Proof} It suffices to show that under any circumstance, the best response of a firm that  {is in permit excess} is lower or equal (in terms of units or permits submitted to the exchange)  {with EC4Ps} than it would be without EC4Ps. Trivially firms that find themselves in permit excess, but which have not change technology, will have the same best responses as before to a given submission level $K_1.$ We assume that the best responses of firm $i$ to $K_1$ are interior (i.e.~they belong to $(0, x^i)$), since otherwise we find boundary solutions where $x^i$ is submitted. Below we write the first order conditions for the interior solutions. The case of no EC4P corresponds to the solution of the equation
\begin{equation}\label{eq:dirvativeNoEC4P}
\exp\left\{\frac{(K_1+x)^2}{(K_1+x)^2-1}\right\}\left(1-\frac{2x(K_1+x)}{((K_1+x)^2-1)^2}\right)=0,
\end{equation}
whereas in the presence of EC4P we must find the root of
\begin{equation}\label{eq:derivativeEC4P}
\exp\left\{\frac{(K_1+x)^2}{(K_1+x)^2-1}\right\}\left(1-\frac{2x(K_1+x)}{((K_1+x)^2-1)^2}\right)=\frac{P_g}{P}.
\end{equation}
Since the mapping $x\mapsto -\frac{2x(K_1+x)}{(K_1+x)^2-1)^2}$ is decreasing, the root of Equation~\eqref{eq:derivativeEC4P} on $(0, x^i)$ is smaller than that of Equation~\eqref{eq:dirvativeNoEC4P}. {By virtue of Lemma~\ref{lm:Higher-price}, we know that any additional permits submitted by firms who are not eligible to cash--4--permits will not restore the total supply to its pre--EC4P levels, which concludes the proof.}

\end{Proof}

\begin{remark}\label{rem:impactEC4Ppayoffs}
By construction, if $i\in s(t+1, h),$ for any $h$ we have that
\be
^{4}\phi^i(t+1, h)\ge\phi^i(t+1, h).
\ee
The equality corresponds to the cases of boundary solutions or firms in permit excess that operate under the old technologies. From Proposition~\ref{prop:greaterProfits}, we also get that if $i\in d(t+1, h)$ then
\be
^{4}\phi^i(t+1, h)\le\phi^i(t+1, h).
\ee
\end{remark}

\subsection{The evolution of the technological vector with EC4P}

In the previous section we have shown that the Nash--games played at the beginning of each period, and which serve to set the expected allowance prices and govern the evolution of the technological vector, are still well defined in the EC4P--setting. It should be kept in mind that the payoff functions $^{4}\Psi^i$ depend on the technology vector $h.$ Furthermore, Proposition~\ref{prop:greaterProfits} indicates that it is in the interest of those firms operating under the new technology to reduce permits availability for exchange purposes. Such a strategy's impact on prices increases the compliance costs when firms are in permit shortage, which in turn affects the evolution of the technological vector. \vspace{0.2cm}

It is not straightforward to compare (in general) the evolution of the technological vectors $h(t)$ and $^{4}h(t),$ under identical primitives and identical triples $\big(T, \{N^i(t)\}_{i\in\I}, P\big).$ The caveat is that, depending on the allocation schedule and the emissions processes, firms may switch back and forth from being on the supply--side to being on the demand--side of the market. This results in the following phenomenon: over some periods the introduction of EC4Ps benefit a certain firm, whereas the latter might be worse off over other periods (in comparison to its position had the EC4Ps not been introduced).\vspace{0.2cm}

The regulator must take into account these dynamics when introducing the EC4P into the policy. There is a delicate balance between the policy level ($\{N^i(t)\}_{i\in\I},$ $P, P_g$) and the cost for adopting the low pollution--emitting technology. As we show in  Section~\ref{ssec:ExEC4P}, the policy regulator can carefully choose such parameters and control the timing and adoption level of the new technologies.

\subsection{Discussion}\label{ssec:ExEC4P}

In this section we analyze our findings and discuss the dynamic incentives to adopt low pollution--emitting technologies employing a numerical example. Our main point of interest is to investigate how the presence of EC4Ps affects the strategic trading behavior of firms in permit excess and the firms' overall incentives to adopt new technologies. \vspace{0.20cm}

\begin{figure}[ht!]
\begin{tabular}{cc}
\subfigure[\label{fig:Alloc} Firm--wise allocation of permits.]
{\includegraphics[width=0.51\textwidth]{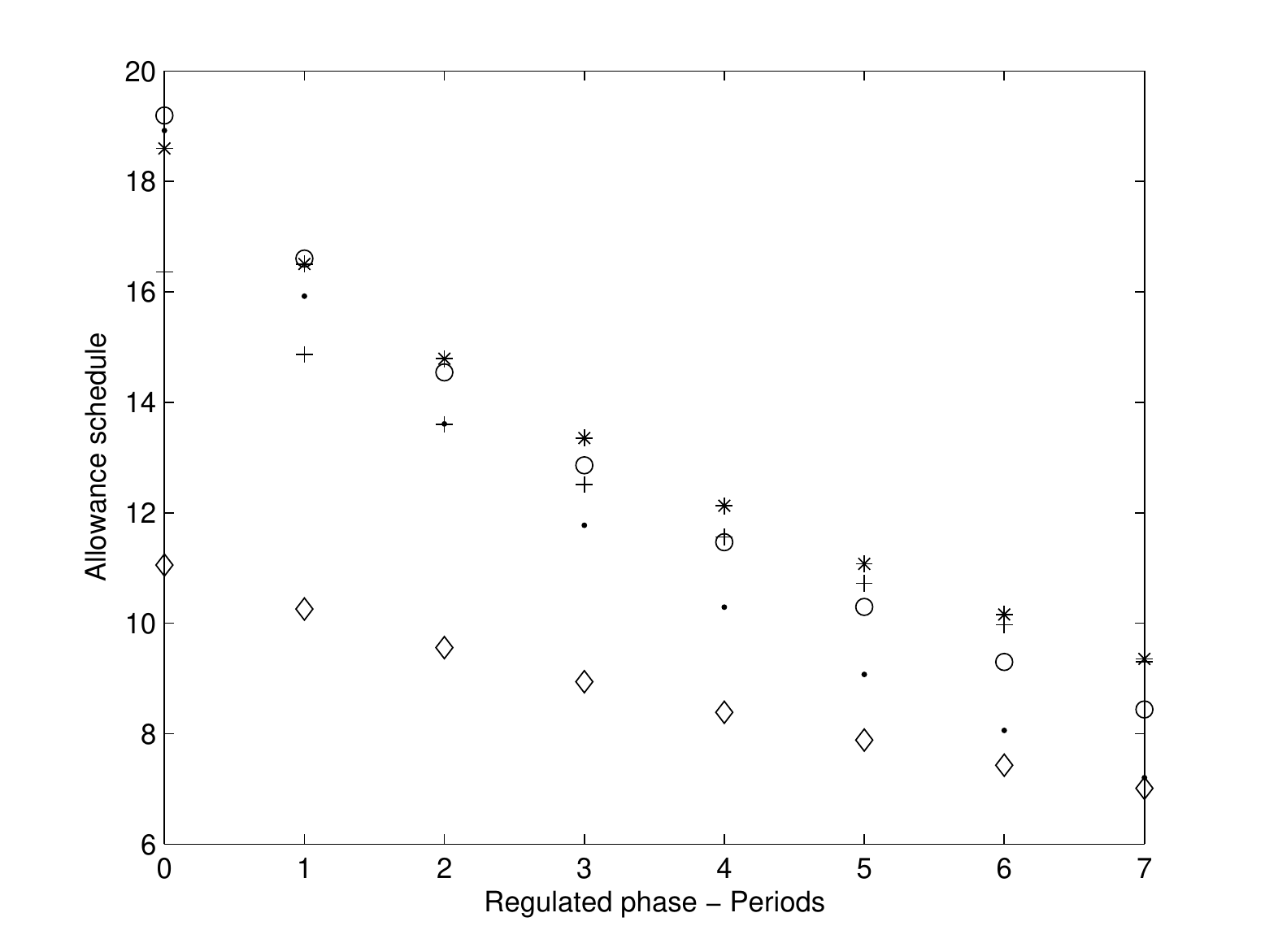}} &
\subfigure[\label{fig:Prices} A possible realization of the permit price.]
{\includegraphics[width=0.47\textwidth]{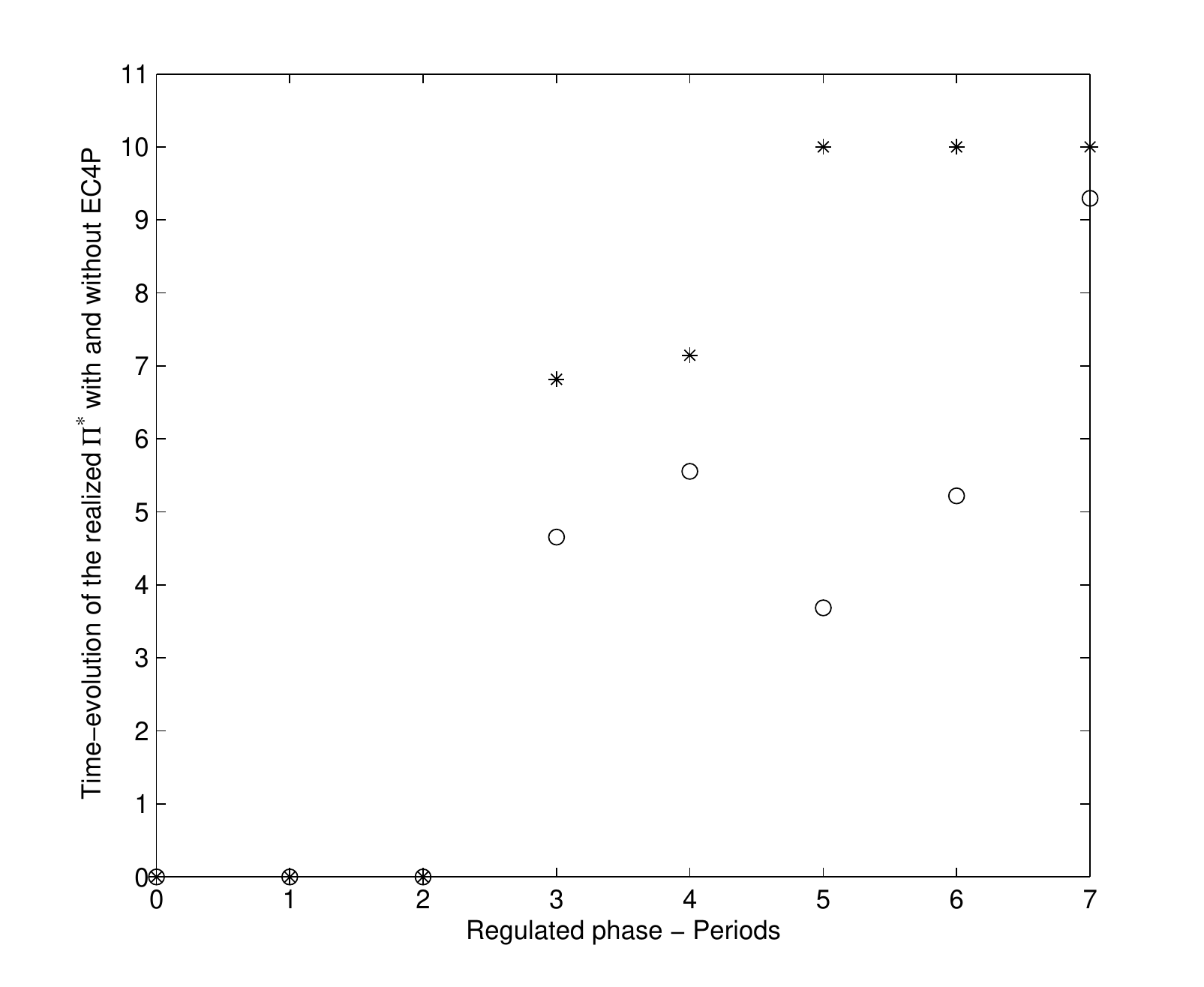}} \\
\end{tabular}
\caption{\label{figure:3} The left diagram represents the unique allocation path of permits to 5 regulated firms. The right diagram represents a possible path of the exchange value of allowances with EC4P (stars) and without EC4P (points).}
\end{figure}

We consider a five--firm, eight--period scenario. The penalty $P$ has been set to 10 and $P_g=5.$ {Later we analyze the effect of different levels of $P_g.$} Recalling the parametric family introduced in Section \ref{sec:regulator}, Figure~\ref{fig:Alloc} shows the allocation schedule for $\alpha\in[-1.5,-0.4]$ and $\beta\in[20,25]$. Figure~\ref{fig:Prices} a possible evolution (a path)\footnote{Recall that viewed from $t=0,$ the allowances price process is a random variable.} of the price process where $u_o \in[1.13,1.15],$ $d_o\in[1.05,1.07],$ $u_n\in[1.08,1.10],$ and $d_n\in[1.02,1.04].$ The initial emission level is the same for each firm, $Q(0)=100,$ the time-constant probability is $q=0.5,$ and the vector-cost to adopt the new technology is $C_n \in[100,80].$ {Having chosen these parameters, the first (last) firm is characterized by higher (lower) emissions and higher (lower) costs for technology adoption.}

\begin{figure}[ht!]
\begin{tabular}{cc}
\subfigure[\label{fig:IndNoEC4P} Firm--wise technological adoption without EC4P]
{\includegraphics[width=0.48\textwidth]{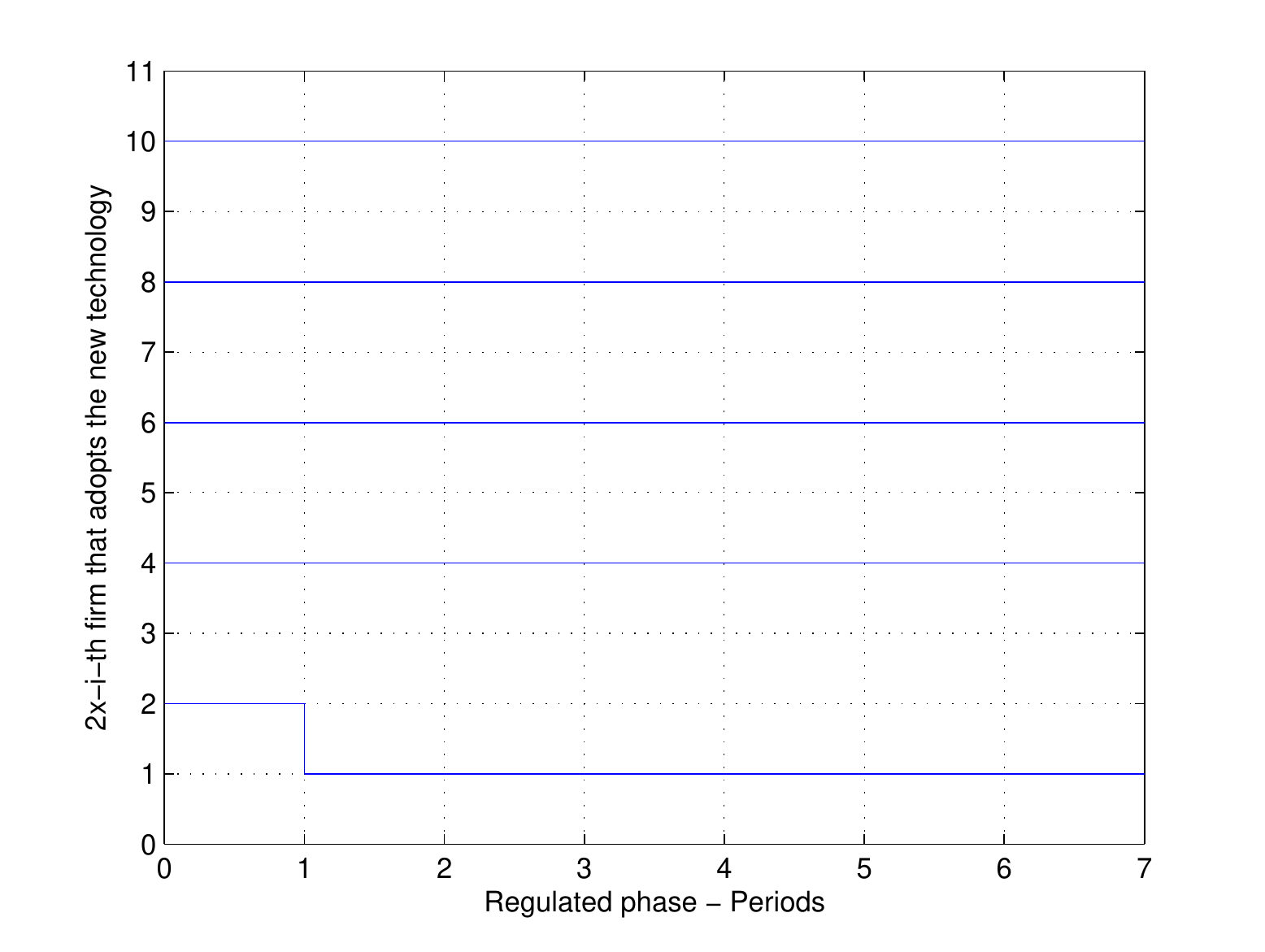}} &
\subfigure[\label{fig:IndYesEC4P}Firm--wise technological adoption with EC4P ]
{\includegraphics[width=0.48\textwidth]{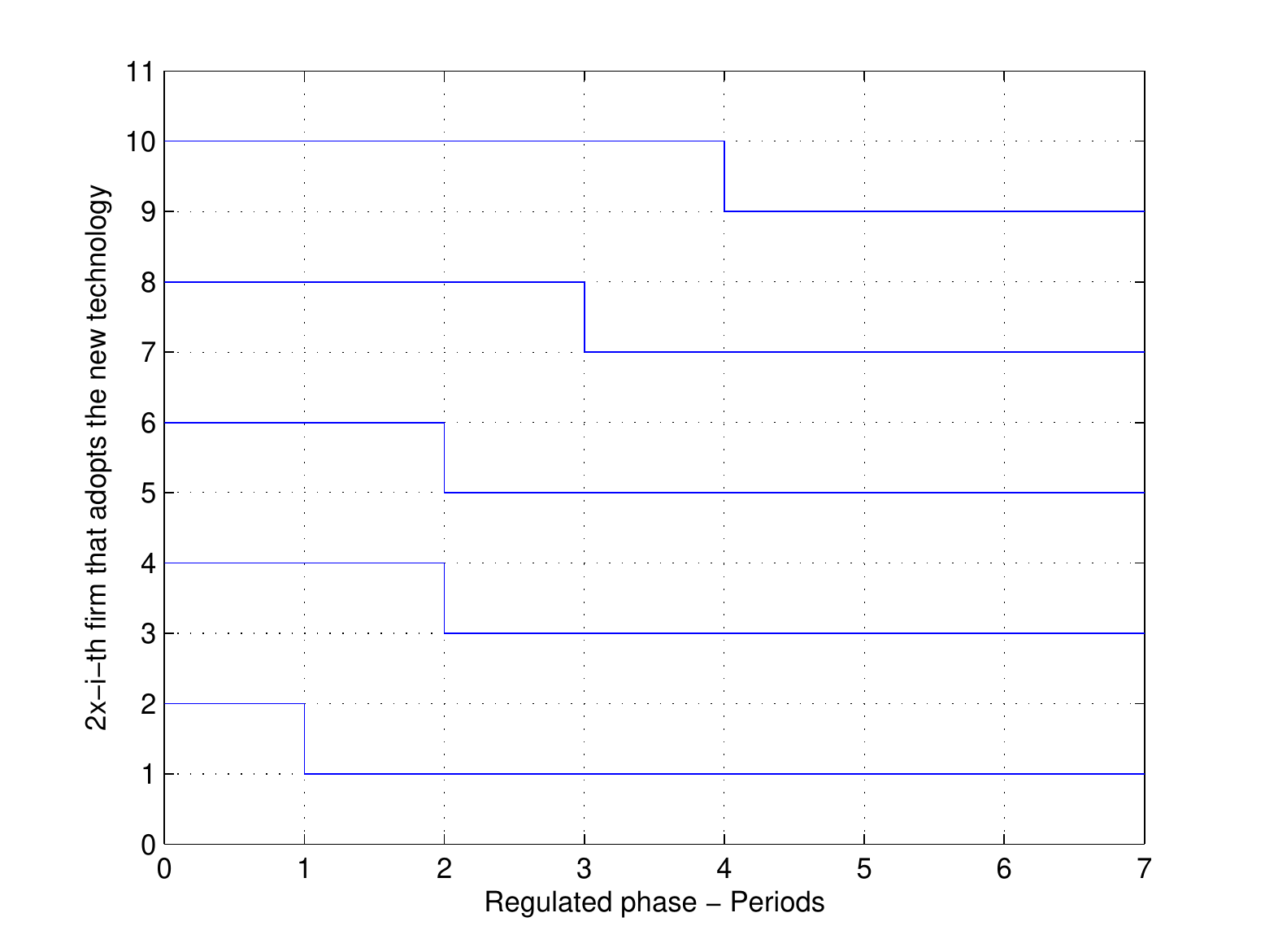}} \\
\end{tabular}
\caption{\label{figure:4} Evolution of the realized firm-specific technology vector without EC4P (left diagram) and with EC4P (right diagram).}
\end{figure}

\begin{figure}[ht!]
\begin{tabular}{cc}
\subfigure[\label{fig:AggNoEC4P} Aggregate technology adoption without EC4P]
{\includegraphics[width=0.48\textwidth]{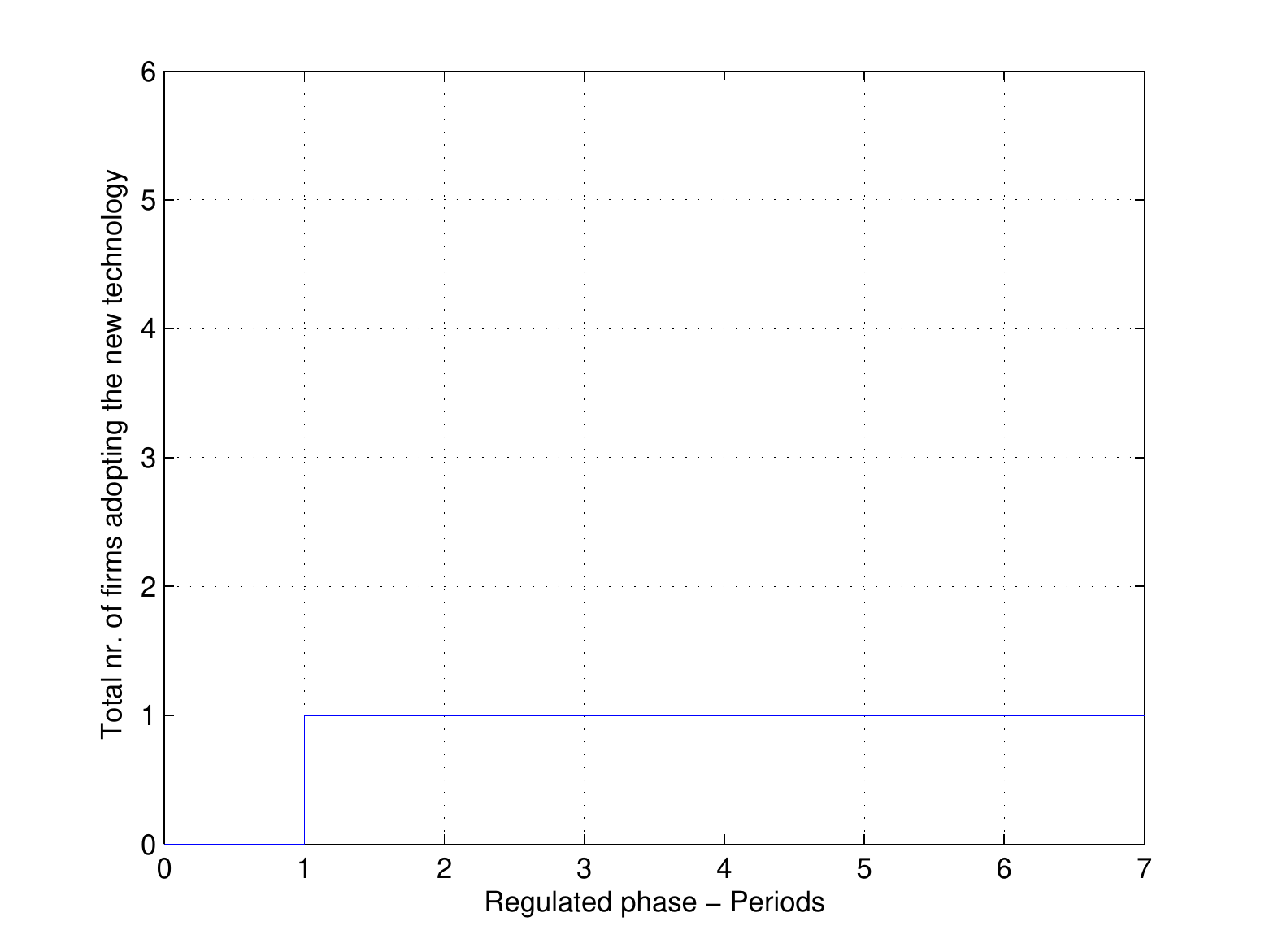}} &
\subfigure[\label{fig:AggYesEC4P} Aggregate technology adoption with EC4P ]
{\includegraphics[width=0.48\textwidth]{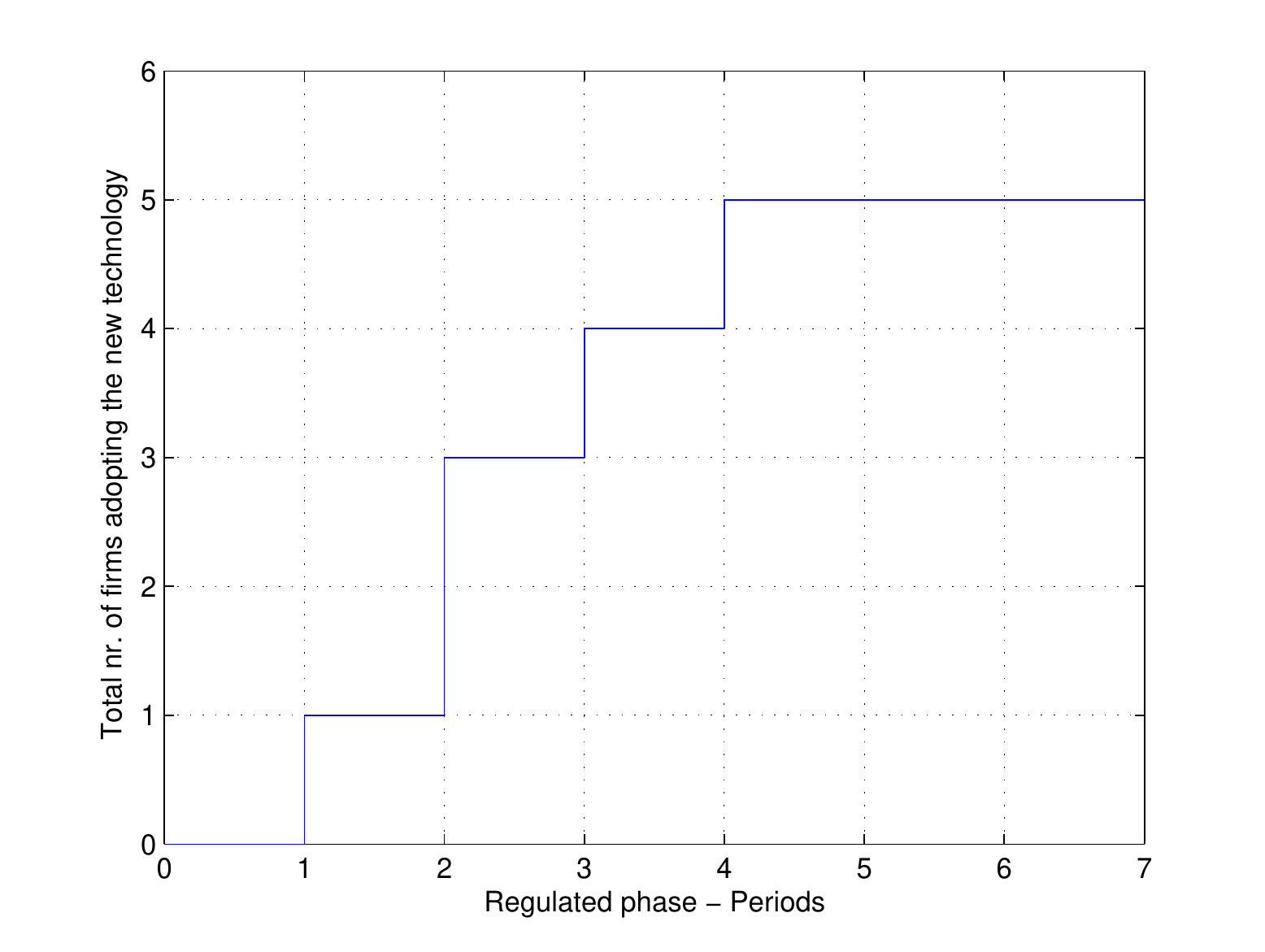}} \\
\end{tabular}
\caption{\label{figure:5} Evolution of the realized firm-specific technology vector in aggregate terms without EC4P (left diagram) and with EC4P (right diagram).}
\end{figure}

\noindent Notice the higher allowance value in Figure~\ref{fig:Prices} in the presence of EC4Ps (stars). Figures~\ref{fig:IndNoEC4P} and~\ref{fig:IndYesEC4P} represent the individual evolution of the technological vector. Here a downwards jump indicates adoption of low pollution--emitting technology.  Similarly, Figures~\ref{fig:AggNoEC4P} and~\ref{fig:AggYesEC4P} show the evolution of the technology vector in aggregate terms. Notice the higher {aggregate level of technology adoption, as expected,} in the presence of EC4Ps. \vspace{0.20cm}

\begin{figure}[ht!]
\begin{tabular}{cc}
\subfigure[\label{fig:Pg35} Aggregate technology adoption for $P_g=3.5$]
{\includegraphics[width=0.45\textwidth]{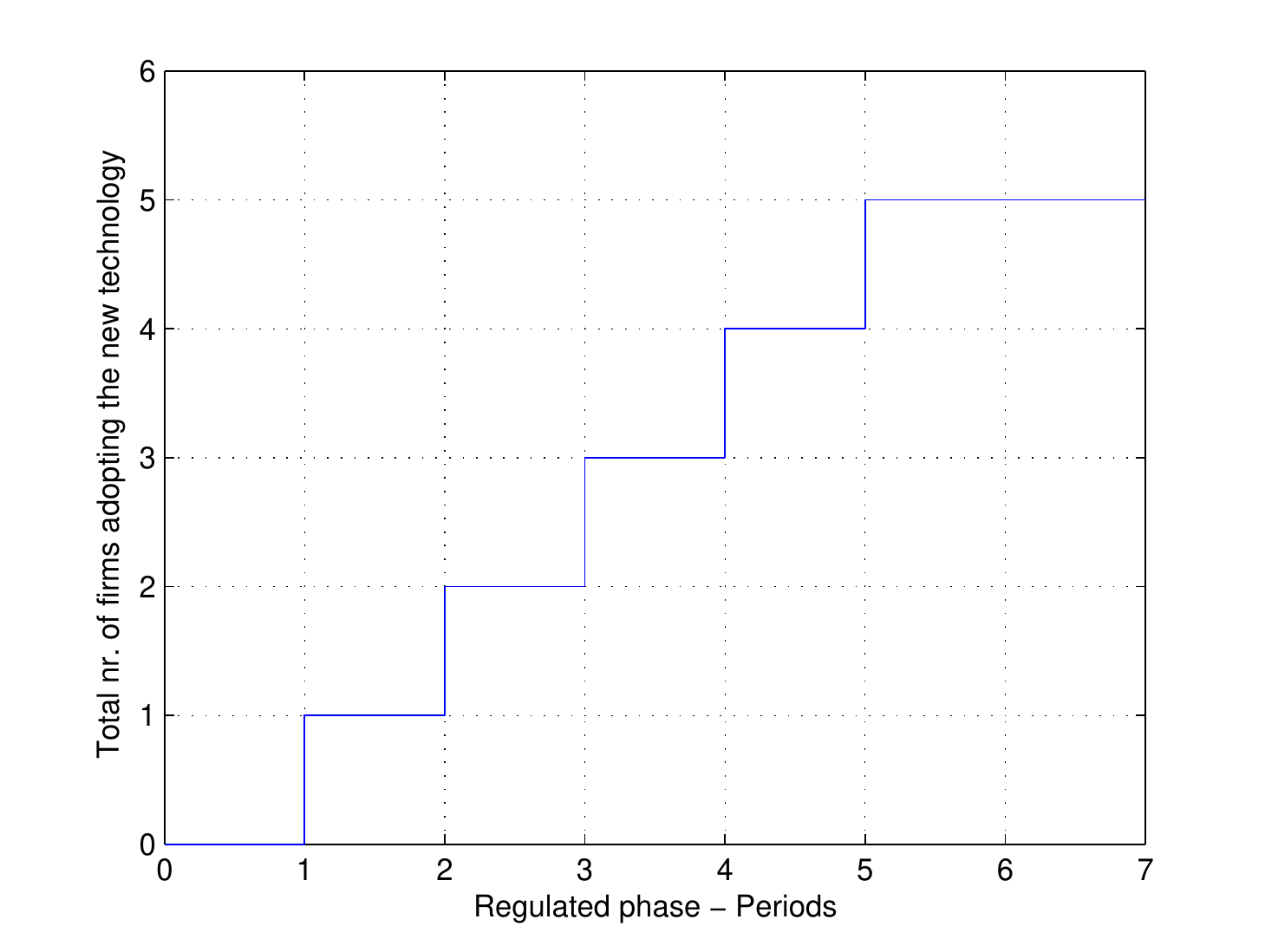}} &
\subfigure[\label{fig:Pg45} Aggregate technology adoption for $P_g=4.5$ ]
{\includegraphics[width=0.45\textwidth]{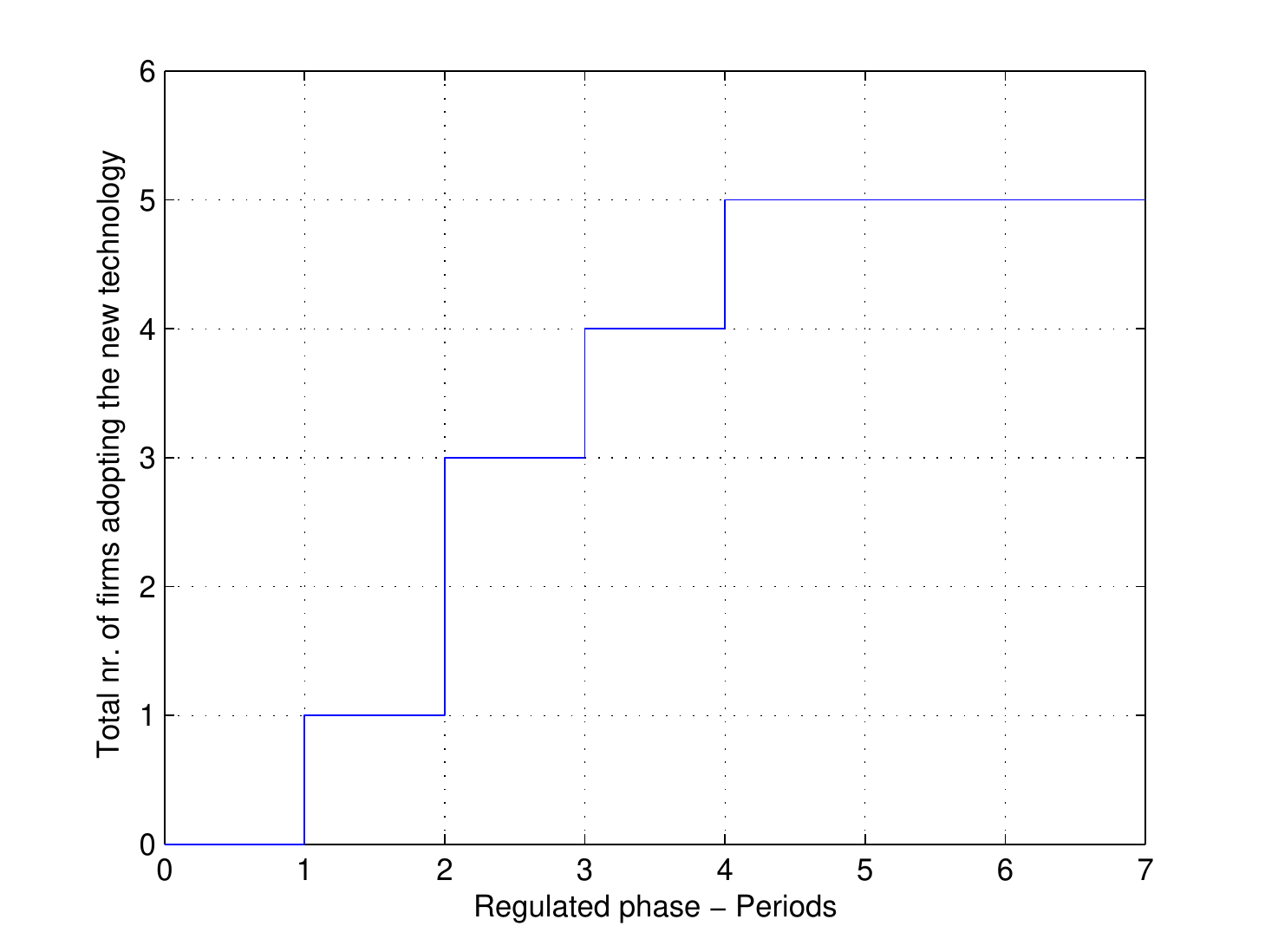}} \\
\end{tabular}
\caption{\label{figure:6} Evolution of the realized technology vector in aggregate terms with different levels of $P_g.$}
\end{figure}

\newpage
\noindent Next we look at the impact that different levels of $P_g$ have on the  {level of technology adoption.} In our example we keep the same parameters as above, save for $P_g,$ which varies between $1.5$ and $4.5.$ It does not come as a surprise that the higher the level of the policy, $P_g,$  the higher the aggregate level of firms adopting new technologies. What is interesting to observe is that by controlling the policy level, $(P, P_g),$ the regulator is also able to affect the timing of the technology adoption. Figure \ref{fig:Pg45} shows that by increasing price support, from $P_g=3.5$ to $P_g=4.5,$ the regulator can accelerate the adoption of the new technologies. However, this decision would increase the cost of the policy. It is part of regulator's task, therefore, to balance the trade--off between inducing rapid technology adoption and having to pay too high a cost for all the EC4P's that might be cashed. This last problem is further analyzed in Section~\ref{sec:SelfFin} below.

\begin{figure}[ht!]
\begin{tabular}{cc}
\subfigure[\label{fig:Pg15} Aggregate technology adoption for $P_g=1.5$]
{\includegraphics[width=0.45\textwidth]{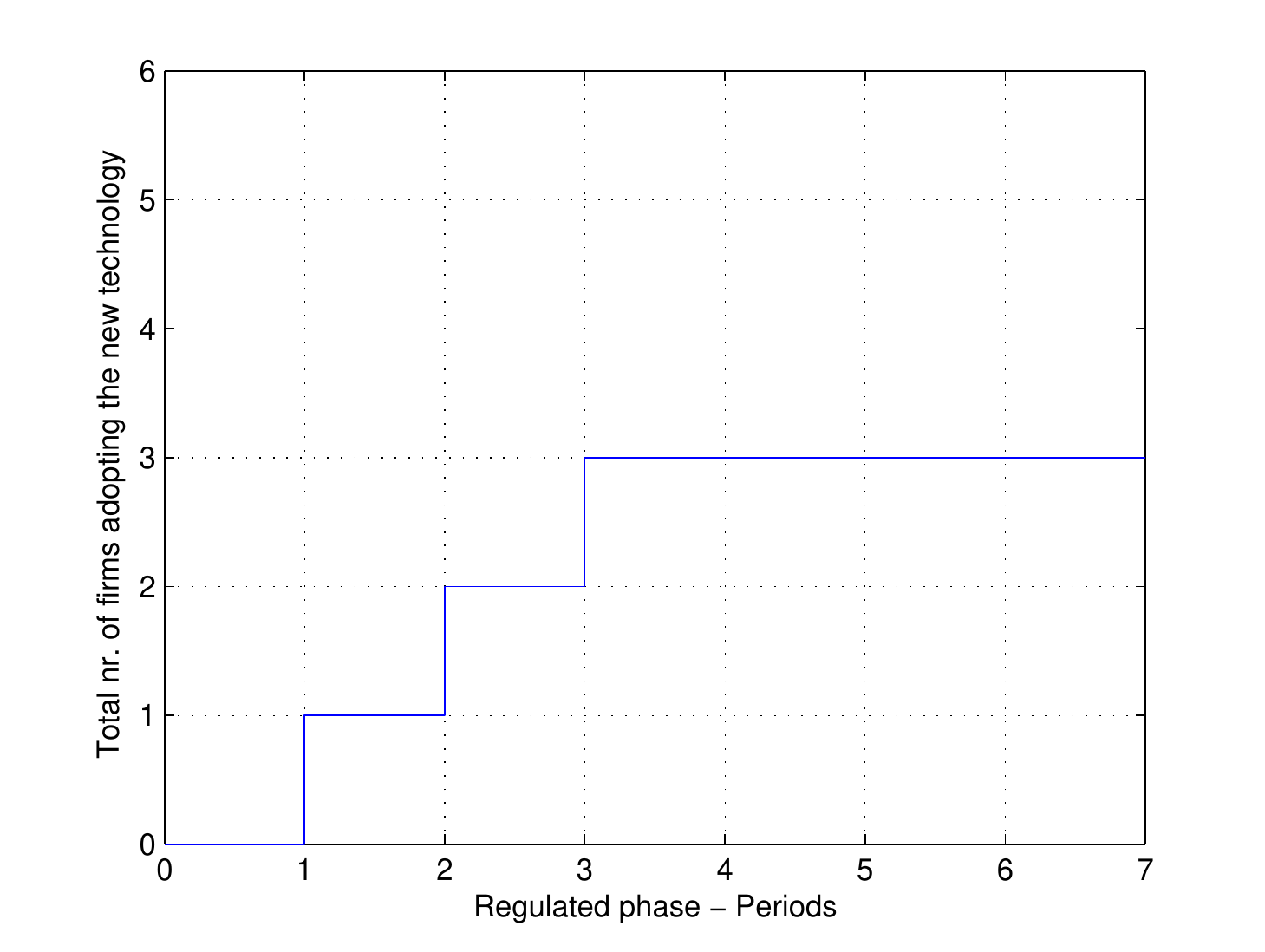}} &
\subfigure[\label{fig:Pg25} Aggregate technology adoption for $P_g=2.5$ ]
{\includegraphics[width=0.45\textwidth]{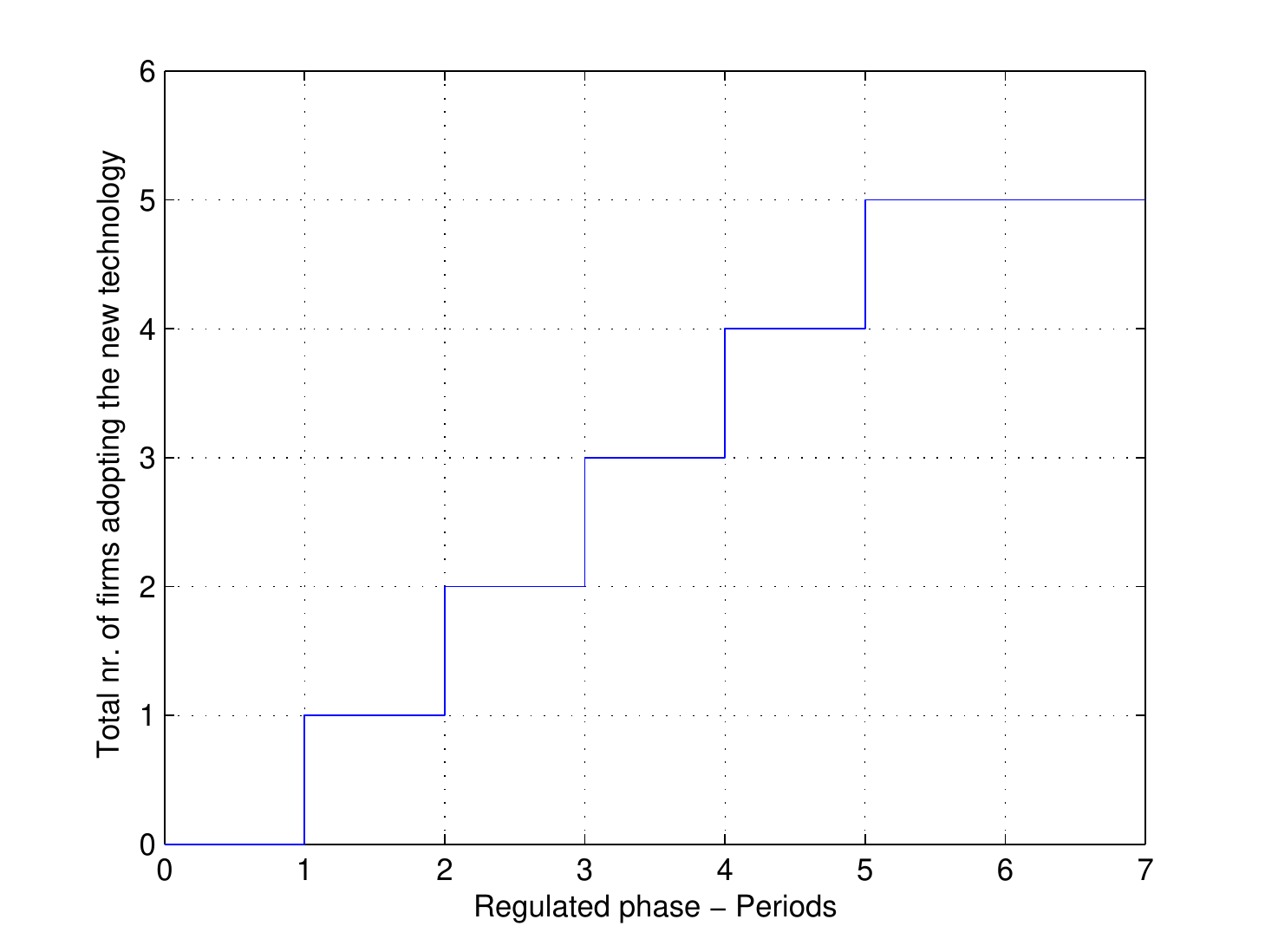}} \\
\end{tabular}
\end{figure}

\newpage
\section{A \textit{Self--financing} Policy with EC4Ps}\label{sec:SelfFin}

In this section we discuss an approach to perform a cost analysis of a policy that includes EC4Ps. The purpose of introducing this additional instrument is to establish a credible policy that dynamically promotes technology adoption. This, however, comes at the expense of the payments to be made to firms that exercise their EC4Ps. Yet, the regulator collects funds from the firms that make penalty payments, which may be used to (partially) cover the cashed EC4Ps.\footnote{We are not considering the costs for running, monitoring and controlling the permit system.} Below we present a methodology to assess the likelihood that the collection of such payments renders the policy \textit{self--financing}, i.e.~that tax--payers' funds are not required to cover the cost of its implementation. We use  \textit{translation--invariant risk measures} (an example of which is value--at--risk) to perform an analysis of how plausible it is that a policy turns out to be self--financing. A summary of important properties of risk measures  is provided in Appendix~\ref{app:RM}.\vspace{0.20cm}

The first task at hand is to specify the probability space over which the risk measures will be defined. To this end, let
\be
\Omega := \big\{(a_1,\ldots, a_T)\,\mid\, a_i\in\{u, d\} \big\}
\ee
be the space of paths of the state of the economy over  $[0,T]$. For example, $(u, d,\ldots, d)\in\Omega$ denotes the scenario where the economy is in ``up'' mode over $[0, 1],$ and then remains in ``down'' mode over $(1, T].$ We pair $\Omega$ with the power $\sigma$--algebra $2^{\Omega},$ which consists of all subsets of $\Omega.$ Since we are working under the assumption that states of the economy at different dates are independent (i.e.~``up'' today has no impact on either ``up'' or ``down'' tomorrow), the probability density $(q(t), 1-q(t))$ used by the regulator naturally induces a \textit{reference measure} $\mathbb{Q}$ on $\Omega$ via
\be
\mathbb{Q}\{(a_1,\ldots a_T)\big\}=q_1\cdots q_T,
\ee
where
\be
\begin{array}{cc}
  q_i= & \left\{
         \begin{array}{ll}
           q(t), & \hbox{if}\; a_1=u(t) \\
           1-q(t), & \hbox{otherwise.}
         \end{array}
       \right.
\end{array}
\ee
In fact, if $q(t)\equiv q,$ then $\mathbb{Q}\{(a_1,\ldots a_T)\big\}=q^n (1-q)^{T-n},$ where $n$ is the number of ``ups'' and, consequently, $T-n$ is the number of ``downs''. Let $X_I$ and $X_0$ be, respectively, the amount of money the regulator collects from penalty payments and the aggregate payments made to EC4P holders during the phase. Notice that for each $\omega\in\Omega,$ $X_I(\omega)$ and $X_O(\omega)$ are well defined via the market mechanisms described in the previous sections. Moreover, by construction $X_I, X_O\in\Ll^{\infty}(\Omega, 2^{\Omega}, \mathbb{Q}).$  \vspace{0.2cm}

Recall that a \textit{law invariant} (convex) risk measure is a (convex) mapping
$\rho:\Ll^{\infty}(\Omega, 2^{\Omega}, \mathbb{Q})\to\re$ such that $\rho(X)=\rho(Y)$ holds for any two $X, Y$ in $\Ll^{\infty}(\Omega, 2^{\Omega}, \mathbb{Q})$ that have the same distribution under $\mathbb{Q}.$ From this point on we assume that choices of risk measures are made from the family of law--invariant ones. For a given $\rho,$  the assessment $\rho(X_I-X_O)$ provides a measurement of whether or not the permits system endowed with EC4Ps will be self--financing; namely, if $\rho(X_I-X_O) \le 0,$ then the mechanism is deemed to be acceptable. An important observation is that for fixed $\big\{T, \{N(t)\}\big\},$ the parameters $P$ and $P_g$ completely determine $X_I-X_O,$ and therefore
\be
\rho(X_I-X_O)=:f(P, P_g).
\ee
The function $f(\cdot, \cdot)$ provides a measure of the losses the regulator might face when he chooses the penalty level $P$ and the minimum price guarantee $P_g$ for the EC4P.\vspace{0.2cm}

The distribution of the random variable $X_I-X_O$ depends both on the evolution of the technological vector and the expected emissions. Since decisions regarding technology adoption are made on the grounds of expected positions, the evolution of $h$ is independent of the realizations of the states of the economy. It does not, however,  depend exclusively on the expected (emissions and permits) positions. In fact, the choice of $P$ and $P_g$ has a stark influence on the evolution of $h$ and, implicitly, on the future expected positions. In other words, the dynamic incentive to adopt the new technology, and therefore how the technology vector evolves, depends (from the firms' point of view) on the potential extra profits (a function of unused permits and $P_g$) and avoided costs (a function of non--offset emissions and $P$).\vspace{0.2cm}

Below we use value--at--risk and average value--at--risk to analyze the examples presented in Section~\ref{ssec:ExEC4P} for different levels of $P_g.$  To this end we have performed 2000 Monte Carlo simulations to approximate the cumulative distribution functions (CDFs) and probability distribution functions (PDFs) of $X_I-X_O.$ Since we are working with  law--invariant risk measures, these distribution functions are sufficient to (numerically) compute $\rho(X_I-X_O).$ In Table \ref{tab:valueatrisk} we report the $V@R_{\lambda}(X_I-X_O)$ and $AV@R_{\lambda}(X_I-X_O)$ for some standard confidence levels $\lambda.$

\vspace{0.5cm}
\begin{table}
\centering
\begin{tabular}{ccccc ccccc}
\hline
    \multicolumn{5}{c}{$\rule[-1mm]{0mm}{5mm} V@R_{\lambda}(X_I-X_O)$}
& \multicolumn{5}{c}{$\rule[-1mm]{0mm}{5mm} AV@R_{\lambda}(X_I-X_O)/100$} \\
\cline{1-5} \cline{7-10}
    \multicolumn{4}{r}{$ P_g \hspace{1.1cm}$}
& \multicolumn{5}{c}{$ \hspace{3.0cm} P_g  $} \\
\cline{2-5} \cline{7-10}
$\lambda$ & 1.5 & 2.5 & 3.5 & 4.5 &  & 1.5 & 2.5 & 3.5 & 4.5\\
\hline
0.10            & $-30.8$         & $-274.6$      &  $-397.0$    & $-704.5$ &
                    & $955$           & $137$           &  $-231$       & $-5034$ \\
0.05            &   $-123.5$    &   $-344.6$     &  $-476.3$    & $-798.1$ &
		 & $468$           & $65$             &  $-217$        & $-5315$ \\
0.01            &   $-187.5$    &   $-424.9$     &  $-594.9$    & $-944.7$  &
		 &   $74$           &   $15$            &  $-208$       & $-5557$ \\
\hline
\end{tabular}
\caption{$V@R_{\lambda}(X_I-X_O)$ and $AV@R_{\lambda}(X_I-X_O)$ for standard confidence levels $\lambda=\{0.10;0.05;0.01\}.$}
\label{tab:valueatrisk}
\end{table}

\begin{figure}[ht!]
\begin{tabular}{cc}
\subfigure[\label{fig:CDF15} CDF for $P_g=1.5$]
{\includegraphics[width=0.45\textwidth]{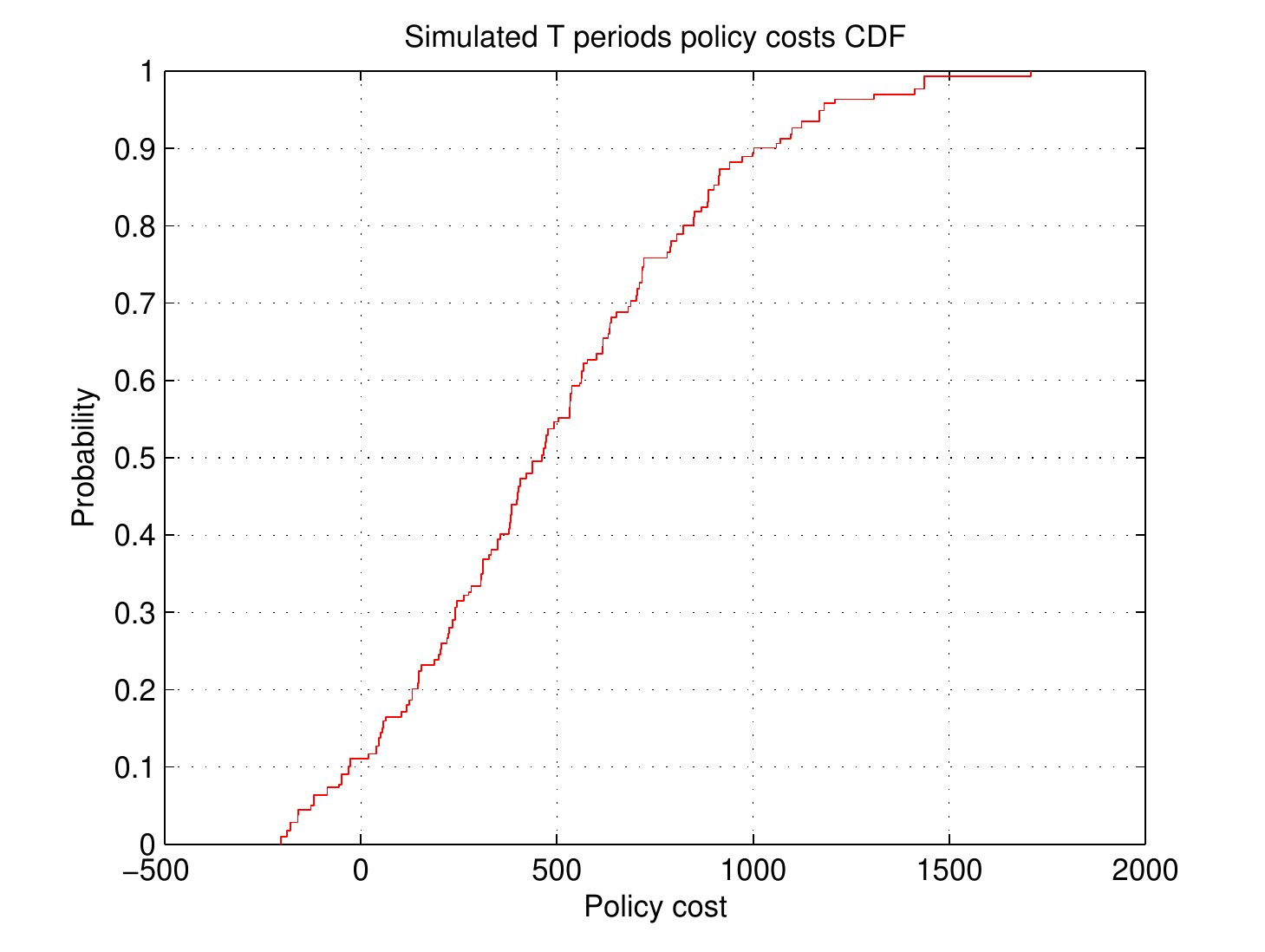}} &
\subfigure[\label{fig:CDF25} CDF for $P_g=2.5$ ]
{\includegraphics[width=0.45\textwidth]{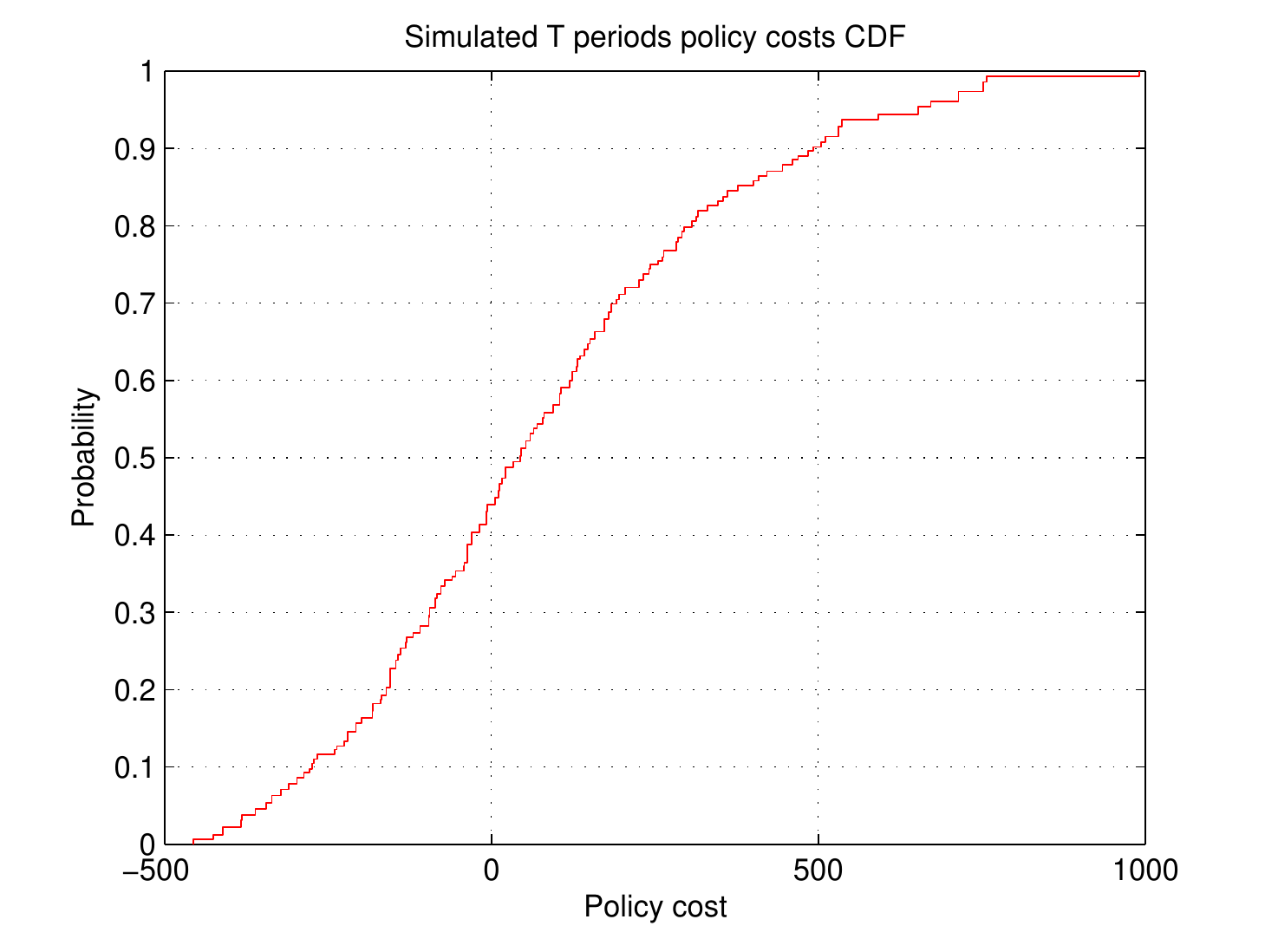}} \\
\end{tabular}
\end{figure}

\begin{figure}[ht!]
\begin{tabular}{cc}
\subfigure[\label{fig:CDF35} CDF for $P_g=3.5$]
{\includegraphics[width=0.45\textwidth]{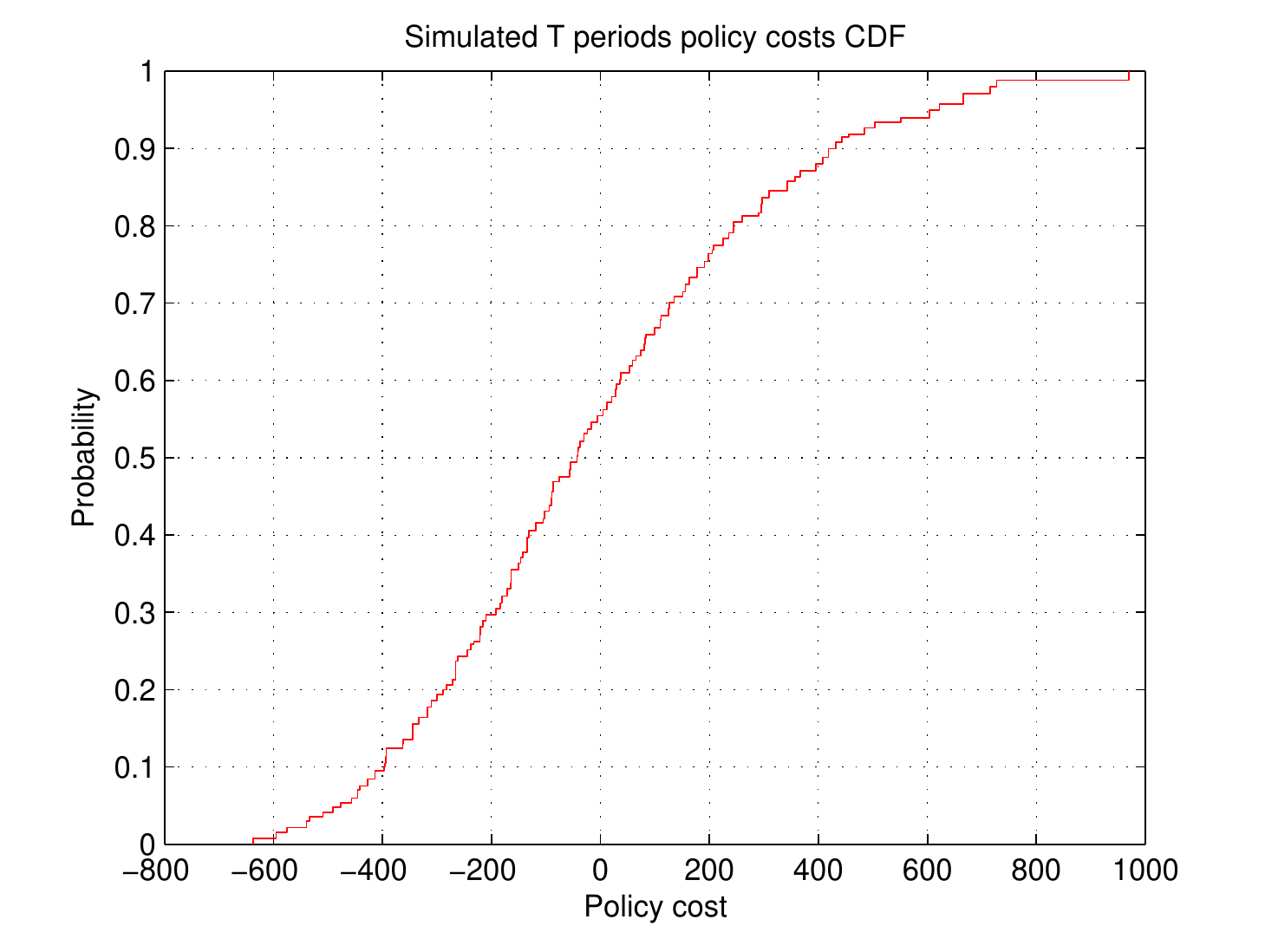}} &
\subfigure[\label{fig:CDF45} CDF for $P_g=4.5$ ]
{\includegraphics[width=0.45\textwidth]{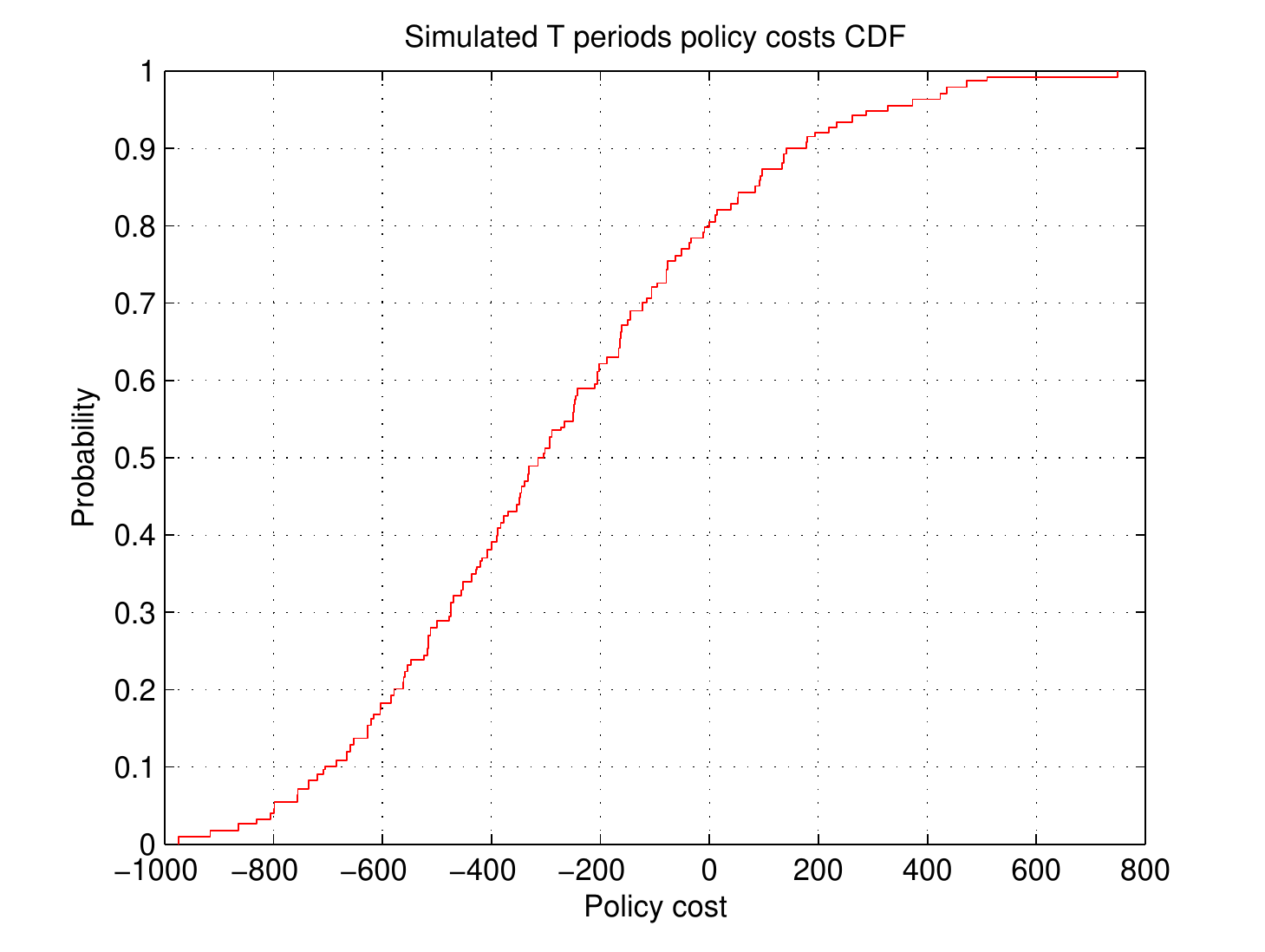}} \\
\end{tabular}
\caption{\label{figure:7} Cumulative distribution functions for different levels of $P_g.$}
\end{figure}

\begin{figure}[ht!]
\begin{tabular}{cc}
\subfigure[\label{fig:PDF15} PDF for $P_g=1.5$]
{\includegraphics[width=0.45\textwidth]{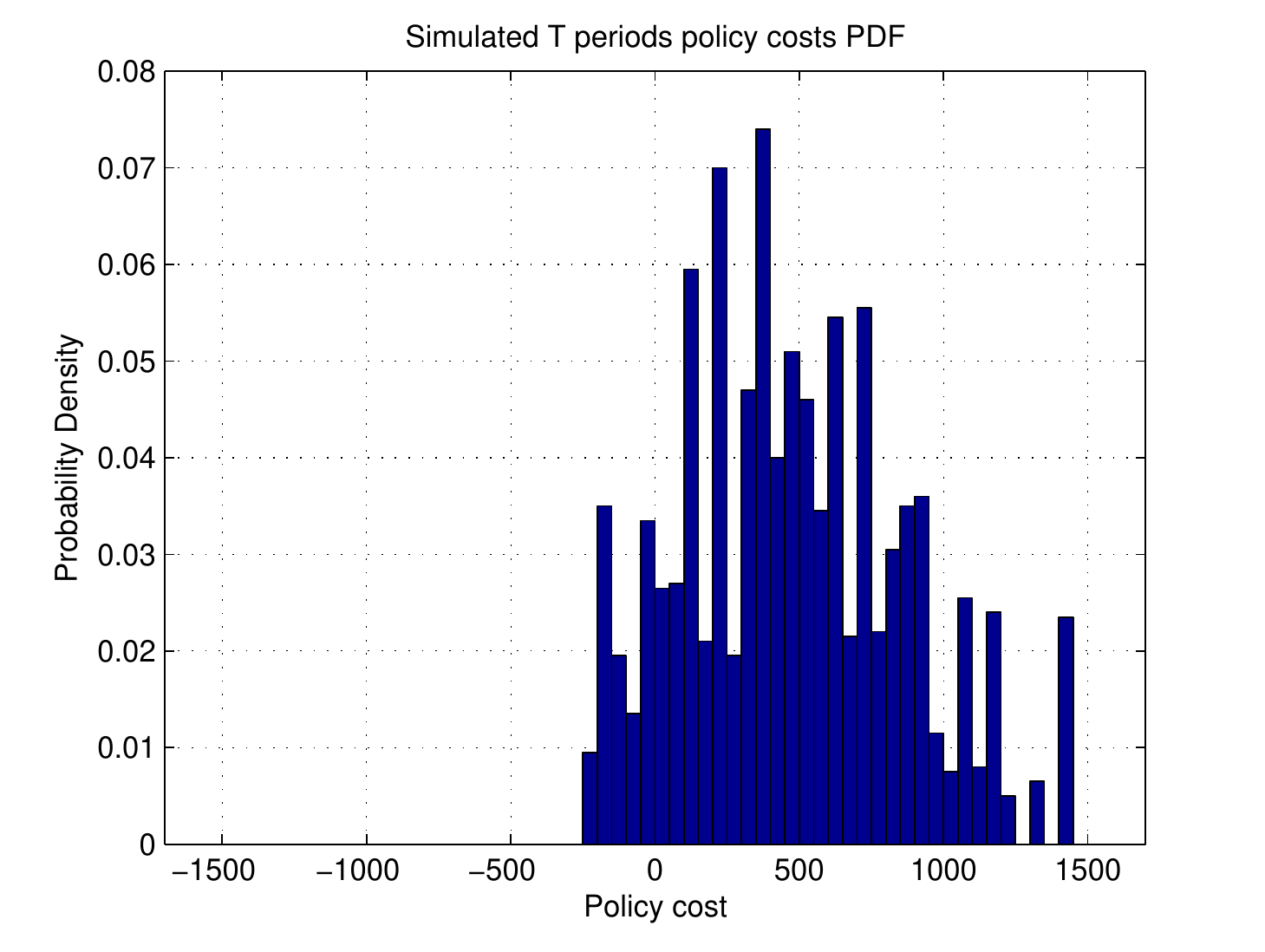}} &
\subfigure[\label{fig:PDF25} PDF for $P_g=2.5$ ]
{\includegraphics[width=0.45\textwidth]{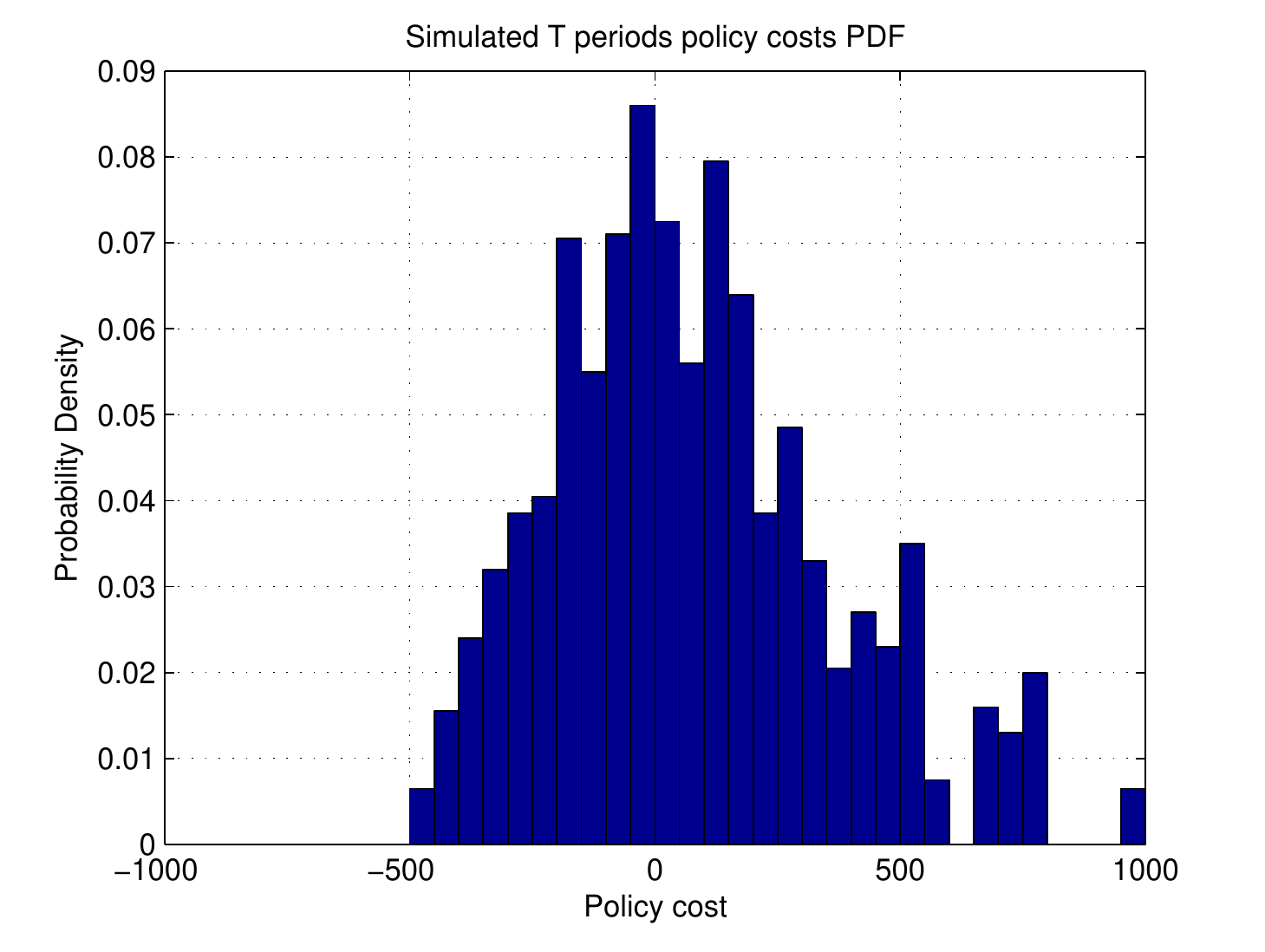}} \\
\end{tabular}
\end{figure}

\begin{figure}[ht!]
\begin{tabular}{cc}
\subfigure[\label{fig:PDF35} PDF for $P_g=3.5$]
{\includegraphics[width=0.45\textwidth]{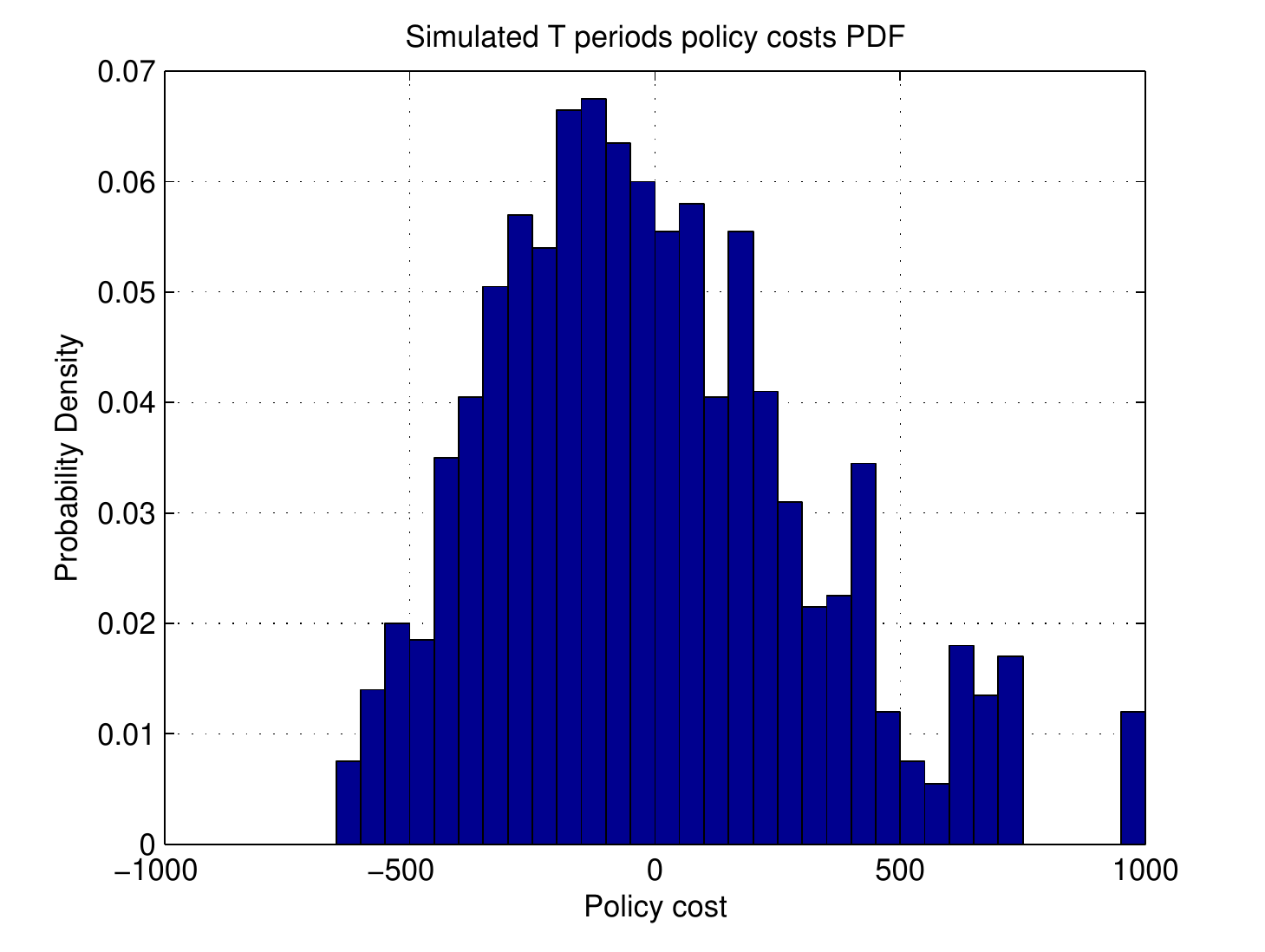}} &
\subfigure[\label{fig:PDF45} PDF for $P_g=4.5$ ]
{\includegraphics[width=0.45\textwidth]{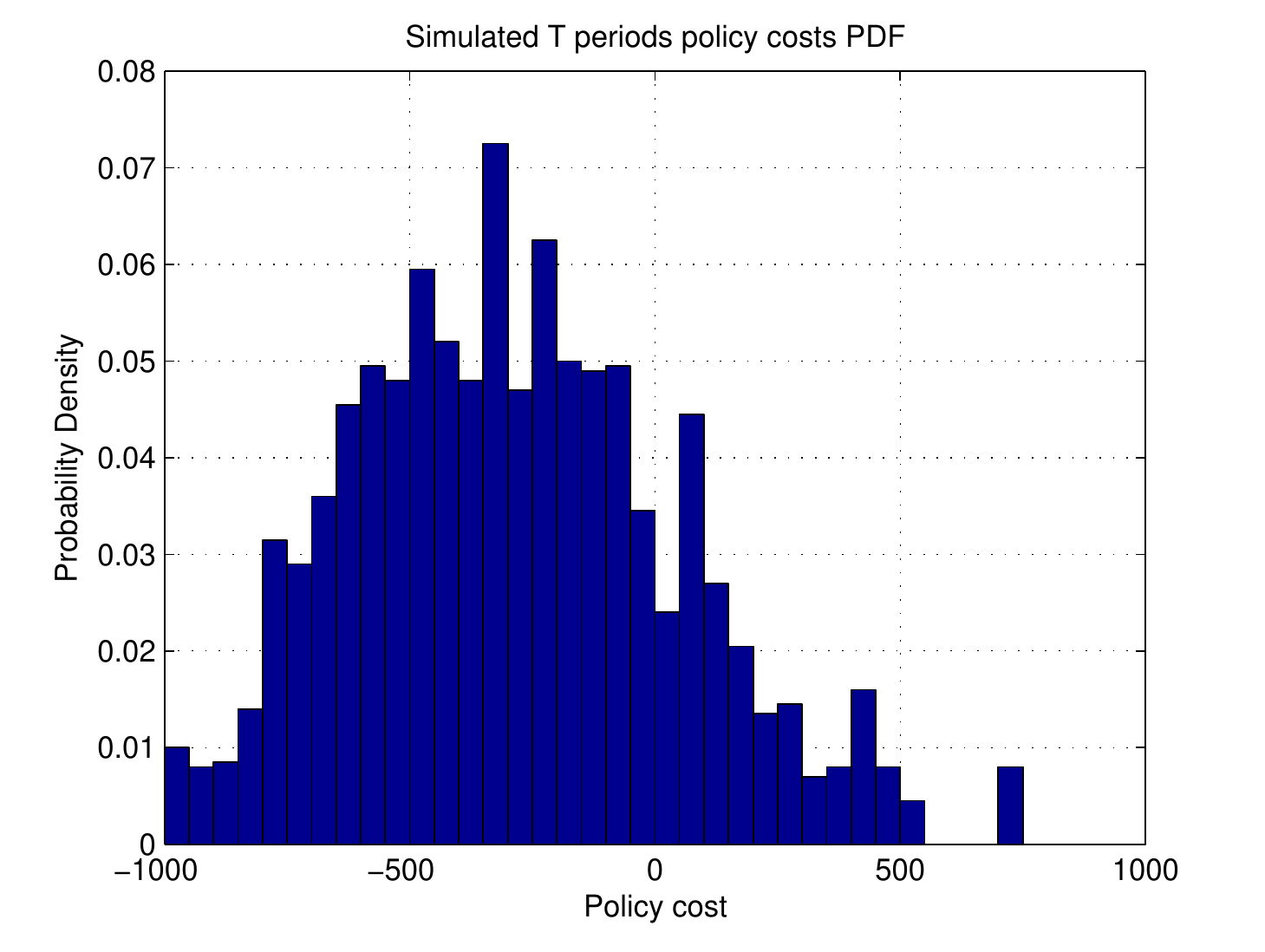}} \\
\end{tabular}
\caption{\label{figure:8} Probability density functions for different levels of $P_g.$}
\end{figure}

\begin{remark}
As it has been pointed out previously, our main aim in this paper is to study the influence of the regulator's decisions on the {dynamic} evolution of the technological vector. On this same token, the introduction of $\rho$ as a tool to measure the likelihood of the policy being self--financing is not done in the spirit of minimizing  {social} costs. In the numerical implementations in Section~\ref{ssec:ExEC4P}, we compare the effect of different levels of $P$ and $P_g$ on the distribution of $X_I-X_O$ (and therefore on $\rho$), and simultaneously on the adoption of low pollution--emitting technologies. It might be the case that a choice of primitives that yields very rapid  {technology adoption} of all firms  is too socially costly. {The analysis of such scenario is not undertaken here}.
\end{remark}

\begin{remark}
A feature of a self--financing policy is that the payments made to firms that have adopted new technologies via cashed EC4P titles are financed by those that have not (adopted). In layman's terms, the ``dirty'' firms subsidize the ``clean'' ones.
\end{remark}

\newpage
\section{Conclusions}

The value of allowances, and more generally the design of the policy, typically determines the incentive of regulated companies to invest in new technologies or adopt alternative low pollution--emitting technologies. Given that such investment costs are in part sunk, it is important for regulated companies to correctly foresee future allowance allocations and penalties on non-offset emissions so as to plan their compliance strategy. \vspace{0.25cm}

In the first part of this work we have studied how, in the presence of imperfect competition, a system based on transferable permits affects incentives to reduce emissions. In the model the allowance value is generated endogenously and reflects the current level of uncovered pollution (demand), the current level of unused allowances (supply), and the current level of adoption of the low emitting--technologies (the technology vector $h$). Since companies are not assumed to be price--takers, it may very well be in the sellers' best interests to reduce the availability of permits and, consequently, increase the allowance exchange value. Thus, strategic trading behaviors are accounted for in the model. In particular, we have constructed a non-cooperative permit trading game and we have shown it possesses a pure-strategy Nash equilibrium. Furthermore, the adoption profiles of new technologies are also generated endogenously. Technology adoption depends on unanticipated macroeconomic shocks and the future values of allowances, the latter being a function of future permits supply and demand. Assuming that the regulator does not anticipate the adoption of new technology, and that he has made a commitment to the type, level and length of his policy instruments ($N,$ $P$ and $T$), we have investigated the number of regulated firms that adopt low pollution--emitting technologies, as well as the timing of such adoptions. \vspace{0.25cm}

Under transferable permits, it would be desirable for the regulator to intervene in the market in response to economic shocks or the occurrence of technology adoption. By adjusting the level of the policy, the regulator would attempt to restore the dynamic incentive to invest. In the second part of the paper we have proposed the introduction of a policy instrument that attempts to preserve the credibility of the policy, ideally reducing the need for intervention by the regulator. In the spirit of \cite{LT:96b}  and \cite{BHQ:95}, we have introduced a price support instrument that is offered by the regulator to firms that have adopted the new technologies. Again, we have constructed the corresponding non-cooperative permit trading game, we have shown it possesses a pure-strategy Nash equilibrium, and we have proved that this new policy instrument creates a floating price floor that can be considered as a (floating) minimum price guarantee. Echoing the conclusions of \cite{LT:96b}, we have shown that \textit{European--Cash--for--permits} provisions, i.e.~the possibility  of cashing unused permits for a fixed amount, have quite attractive properties. Evidently, this policy has a cost. Recalling that the penalty payments generate potential incomes, we have introduced the notion of \textit{self-financing} policies when tax-payers' funds are not required to cover the payments of the \textit{European--Cash--for--permits}. Based on this definition, and using different pairs $(P,P_g),$ we have numerically assessed how likely it is (in terms of convex risk measures) that the regulator will have to access tax-payers' funds, instead of using penalty payments. Numerical findings confirm that by controlling the policy level, $(P,P_g),$ the regulator can also affect the timing of technology adoption by increasing price support, $P_g,$ and, accepting higher policy costs. This prevents delays in the adoption of the new technologies. \vspace{0.25cm}

\noindent The investment technology is modeled here somewhat rigidly (adopt or not adopt the new technology). There are several other situations that could be considered. First, we could allow a more continuous choice in abatement. In such a model this is possible through a costly choice of different levels of abatement. Secondly, investments create purely private benefits. One could investigate the situation where investing in new technologies would generate innovations in abatement technologies which are public goods.

\begin{appendix}

\section{Convex risk measures on $\Ll^{\infty}$.}\label{app:RM}

This appendix contains  some
results on risk measures defined on $\Ll^{\infty}(\Omega, \mathcal{F}, \Prob).$ \footnote{We refer the reader
to section 4.3 in~\cite{kn:fs} for detailed discussions on this topic.} Up to a change in sign, convex risk measures coincide with robust utility functionals.
The latter are an extension to von Neumann--Morgenstern expected--utility functionals, and they were introduced to address issues of agent irrationality, \`a la Allais´ paradox.

\begin{definition}
\begin{rmenumerate}
\item A {\sl{monetary measure of risk}} on $\Ll^{\infty}(\Omega, \mathcal{F}, \Prob)$ is a function
$\rho:\Ll^{\infty}(\Omega, \mathcal{F}, \Prob)\to\re$ such that for all $X,Y \in\Ll^{\infty}(\Omega, \mathcal{F}, \Prob)$ the
following conditions are satisfied:
\begin{itemize}
\item Monotonicity: if $X\le Y$ then $\rho(X)\ge\rho(Y)$.

\item Cash Invariance: if $m\in\re$ then $\rho(X+m)=\varrho(X)-m$.
\end{itemize}
\item A risk measure is called {\sl coherent} if it is convex and homogeneous of degree 1, i.e., if the following two conditions hold:
\begin{itemize}
\item Convexity: for all $\lambda \in [0,1]$ and all positions
$X,Y \in\Ll^{\infty}(\Omega, \mathcal{F}, \Prob)$:
\be
    \rho(\lambda X+(1-\lambda)Y)\le\lambda\rho(X)+(1-\lambda)\rho(Y)
\ee

\item Positive Homogeneity: For all $\lambda \geq 1$
\be
    \rho(\lambda X)=\lambda\rho(X).
\ee
\end{itemize}

\item The risk measure is called law invariant, if
\be
    \rho(X)=\rho(Y)
\ee
for any two random variables $X$ and $Y$ which have the same law.
\item The risk measure $\rho$ on $\Ll^{\infty}(\Omega, \mathcal{F}, \Prob)$ has the {\sl Fatou
property} if for any sequence of random variables $X_1,
X_2,\ldots$ that converges almost surely to a random variable $X$ we
have
\be
    \rho(X)\le\liminf_{n\to\infty} \rho(X_n).
\ee
\end{rmenumerate}
\end{definition}
The intuition behind the property of translation invariance is the following: if $X$ represents an uncertain future position, and if a position is deemed \textit{acceptable} if its image under $\rho$ is non--positive, then $\rho(X)$ is the amount of cash that makes position $X$ acceptable. In other words, $\rho(X)$ is a {\textit{capital requirement}}. \vspace{.2cm}

A risk measure $\varrho$ that has the Fatou property may be represented it via a Legendre--Fenchel transform. Namely, if we define $\mm_1(\Prob)$ to be the set of probability measures on $\Omega$ that are absolutely continuous (w.r.t $\Prob$)
then given $\rho$ there exists a penalty function $\alpha:\mm_1(\Prob)\to\re\cup\{+\infty\}$ such that
\begin{equation}\label{eq:robustRM}
\rho(X)=\sup_{\Lambda\in \mm_1(\Prob)}\big\{\E_{\Prob}[-X]-\alpha(\Lambda)\big\}.
\end{equation}
 The idea behind such representation is the following: although $\Prob$ should reflect the true distribution of events in $\Omega,$ this could prove to be wrong. As a consequence, one moves away from simply evaluating $\E_{\Prob}[X],$ and instead they perform a robust evaluation as in Equation~\eqref{eq:robustRM}. The larger $\alpha(\Lambda)$ is, the less likely that the scenario represented by a given $\Lambda$ is believed to occur. For instance, if $\alpha(\Lambda)=+\infty$ for all $\Lambda\neq\Prob,$ then an agent who assesses risk using the corresponding measure is simply risk--neutral. Any other choice of $\alpha$ denotes a certain degree of risk aversion, which is intuitively represented by the curvature of the graph of the convex functional $\rho.$ Notice that  $\alpha$ fully characterizes $\rho,$ but it needs not be unique.\vspace{.2cm}

\noindent For $Y \in L^{\infty}(\Omega, \mathcal{F}, \Prob),$ the upper quantile function of $Y$ is defined as
\be
q_Y(t):= \sup\big\{l\in\re\,\big|\,\Prob(Y\le l)> t \big\}.
\ee
Given a {\textit{confidence level}} $\lambda\in (0, 1],$ the {\textit{value--at--risk}} at level $\lambda$ of a position $Y$ is
\be
V@R_{\lambda}(Y):=q_Y(\lambda).
\ee
Given $\lambda\in (0,1],$ the \textit{average value--at--risk} of
level $\lambda$ of a position $Y$ is defined as
\be
    AV@R_{\lambda}(Y):=-\frac{1}{\lambda}\int_0^{\lambda} q_Y(t)dt.
\ee
If $Y \in
L^{\infty}(\Omega, \mathcal{F}, \Prob)$, then we have the following characterization
\be
    AV@R_{\lambda}(Y)=\sup_{Q\in{\cal{Q}}_{\lambda}}-\E_Q[Y]
\ee
where
\be
    {\cal{Q}}_{\lambda}=\left\{Q<<P\,\mid\,\frac{dQ}{dP}\le\frac{1}{\lambda}\right\}.
\ee
It turns out to be the case that the Average Value of Risk can be viewed as a basis
for the space of all law--invariant, coherent risk measures with
the Fatou property. More precisely, we have the following result.

\begin{thm}\label{th:avar} The risk measure
$\varrho: L^{\infty}(\Omega, \mathcal{F}, \Prob) \to \re$ is law--invariant, coherent and has the
Fatou Property if and only if $\varrho$ admits a representation of
the following form:
\be
    \varrho(Y)=\sup_{\mu\in M}\left\{\int_0^1
    AV@R_{\lambda}(Y)\mu(d\lambda)-\beta(\mu)\right\}
\ee
where $M$ is a set of probability measures on the unit interval and $\beta$ plays the role of a penalty function.
\end{thm}

\end{appendix}

\bibliographystyle{apalike}
\bibliography{Literature}

\end{document}